\documentclass[fleqn,usenatbib,usedcolumn]{mnras}
\usepackage{mathptmx}

\usepackage[T1]{fontenc}
\usepackage{ae,aecompl}
\usepackage{mathtools}

\usepackage{graphicx}	
\usepackage{amsmath, amssymb}

\date{\today}
  
\def\[{\begin{equation}}
\def\]{\end{equation}}

\newcommand{\ud}{\mathrm{d}}
\newcommand{\set}{\mathrm{\mathbf{X}}}
\newcommand{\WMF}{W}
\newcommand{\hyps}{A} %Hypsometric curve
\newcommand{\oceanfrac}{f_w}

\newcommand{\stdElevation}{\sigma_{h}}

\newcommand{\habfrac}{f_h}
\newcommand{\thetaearth}{\theta_\oplus}

\newcommand{\Vnorm}{S}
\newcommand{\Vw}{V_\mathrm{w}}
\newcommand{\Vb}{V_\mathrm{b}}
\newcommand{\sigmaX}{\sigma_{\Vnorm}}
\newcommand{\Searth}{\Vnorm_\oplus}

\newcommand{\mS}{\mu_\Vnorm}  % Median saturation

\newcommand{\Ho}{\mathcal{H}_0}
\newcommand{\Hs}{\mathcal{H}_1}

\newcommand{\waterconst}{\alpha}
\newcommand{\observer}{O}
\DeclareMathOperator\erf{erf}

\let\oldhref\href
\renewcommand{\href}[2]{\oldhref{#1}{\hbox{#2}}}
 
\title[The Prevalence of Waterworlds]{Bayesian evidence for the prevalence of waterworlds}
\author[F. Simpson]{
Fergus Simpson\thanks{fergus2@icc.ub.edu}
\\
ICC, University of Barcelona (UB-IEEC), Marti i Franques 1, 08028, Barcelona, Spain. \\ 
}

\begin{document}
\label{firstpage}
\pagerange{\pageref{firstpage}--\pageref{lastpage}}
\maketitle

\begin{abstract}   
Should we expect most habitable planets to share the Earth's marbled appearance?  For a planetary surface to boast extensive areas of both land and water, a delicate balance must be struck between the volume of water it retains and the capacity of its perturbations.  These two quantities may show substantial variability across the full spectrum of water-bearing worlds. This would suggest that, barring strong feedback effects, most surfaces are heavily dominated by either water or land.  Why is the Earth so finely poised? To address this question we construct a simple model for the selection bias that would arise within an ensemble of surface conditions. Based on the Earth's ocean coverage of $71\%$, we find substantial evidence (Bayes factor $K \simeq 6$) supporting the hypothesis that anthropic selection effects are at work. Furthermore, due to the Earth's proximity to the waterworld limit, this model predicts that most habitable planets are dominated by oceans  spanning over $90\%$ of their surface area ($95\%$ credible interval). This scenario, in which the Earth has a much greater land area than most habitable planets, is consistent with results from numerical simulations and could help explain the apparently low-mass transition in the mass-radius relation.   
\end{abstract}

\section{Introduction}
\begin{figure*} 
\includegraphics[width=\columnwidth]{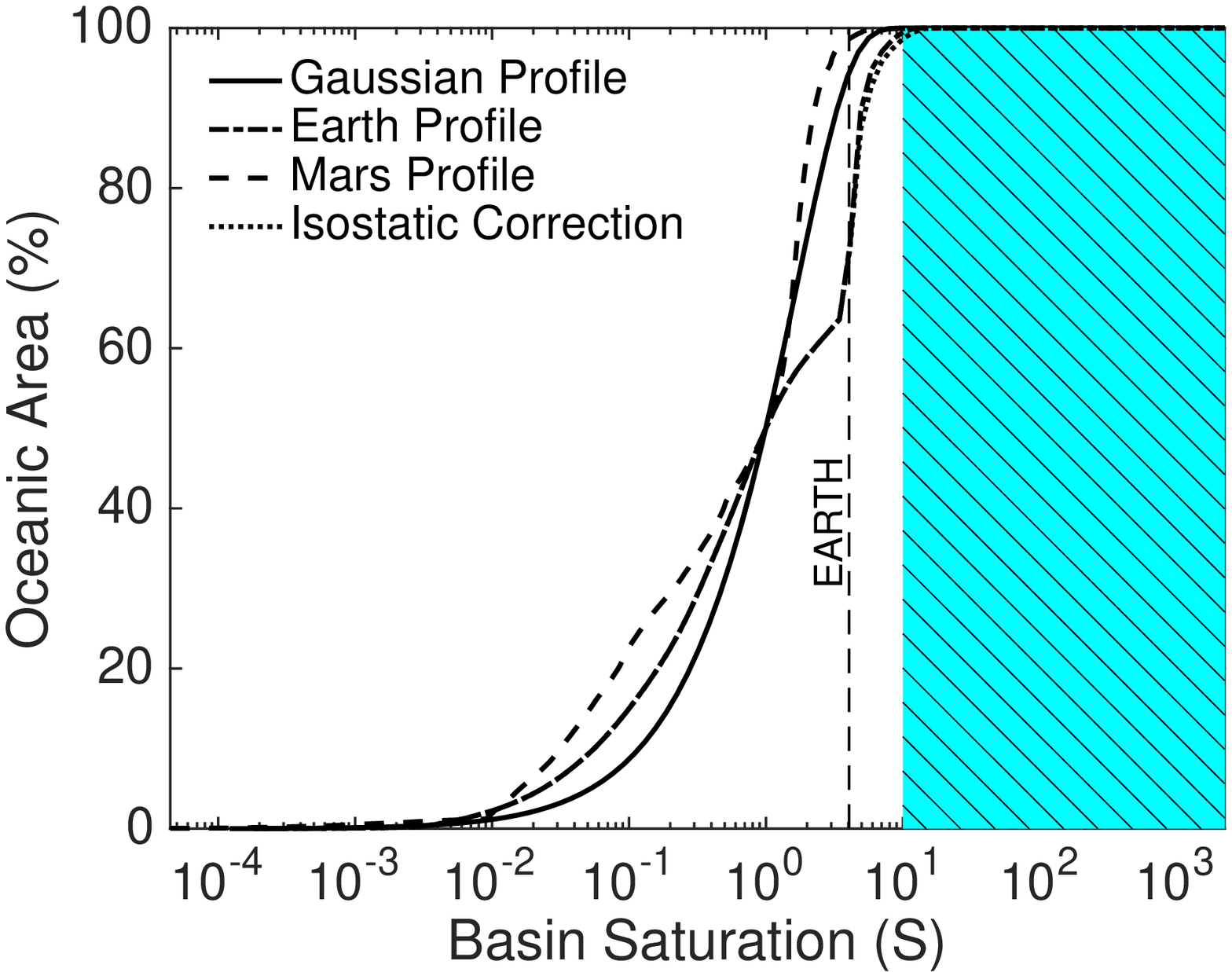} 
\hspace{0.2cm}
\includegraphics[width=\columnwidth]{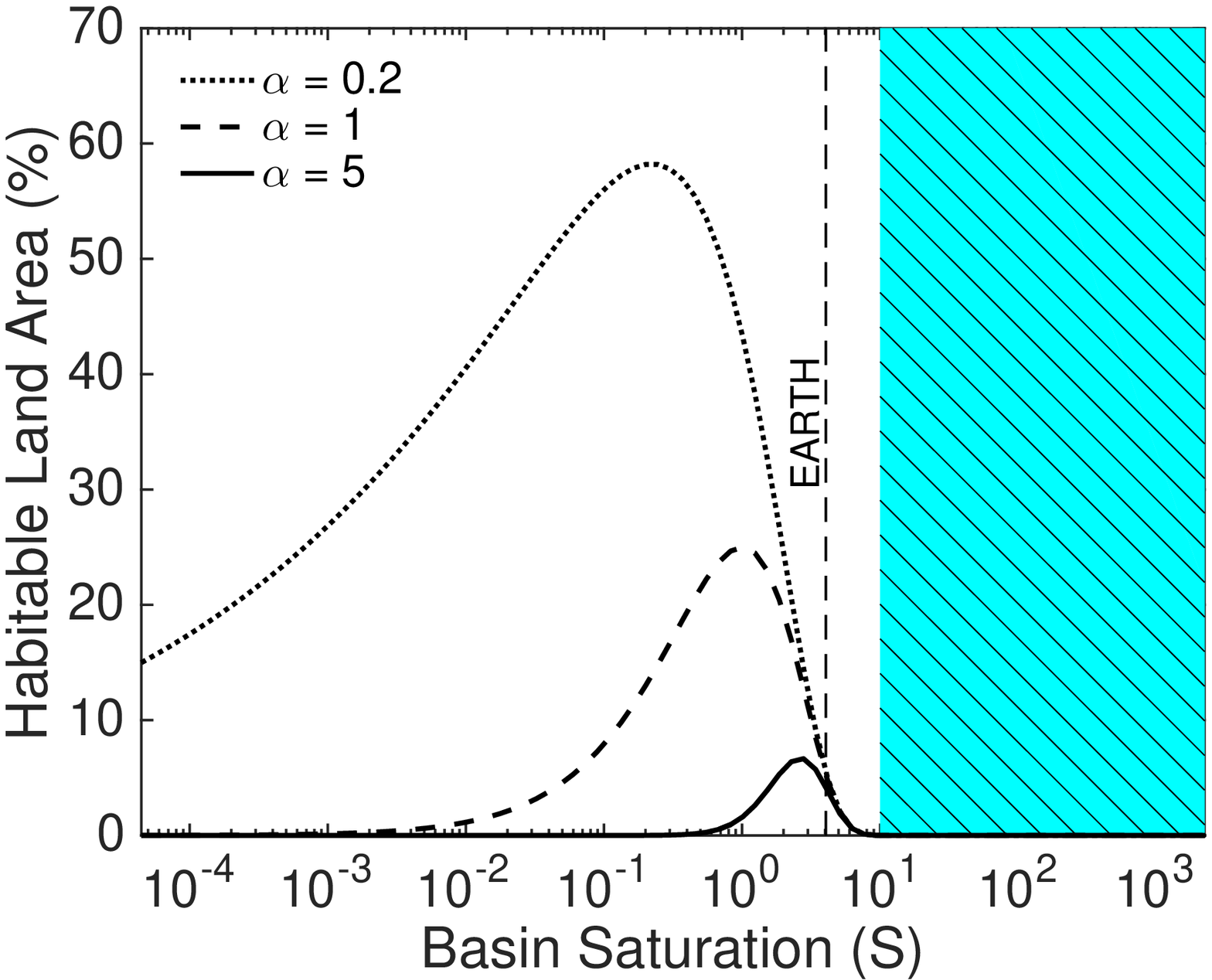} 
\caption{The oceanic fine-tuning problem.  \emph{Left:} The ocean coverage as a function of the surface water volume, normalised in terms of the capacity of surface perturbations. The solid line describes any solid surface with a Gaussian hypsometry. This will serve as our model for the statistical average across all habitable worlds.  The dot-dashed line depicts the effect of adopting the shape of the Earth's elevation profile  \citep{eakins2012hypsographic}, while the dotted line (barely distinguishable) corrects for the isostatic depression of the seabed. The thick dashed line shows the response for the elevation profile of Mars. The thin vertical dashed line demonstrates that the Earth's value of $\Searth \simeq 4$ is precariously close to the waterworld limit. \emph{Right:} Three different models of the habitable land area, expressed as a fraction of the total surface area.  The dotted, dashed and solid lines correspond to values of $\alpha = \frac{1}{5}, 1, 5$ respectively, and the habitable area is defined by equation (\ref{eq:H}). These curves are generated using a Gaussian hypsometry, as depicted in the left hand panel.}
\label{fig:landfrac}
 \end{figure*}
 
The methane seas of Titan are the only exposed bodies of liquid known to exist beyond our planet. They differ markedly from the Earth's oceans, not only in terms of chemistry, but also in their modest expanse.  As a result, Titan is a world whose surface remains heavily dominated by dry land. Remarkably, \citet{dermott1995tidal} were able to deduce this fundamental feature long before detailed surface observations became available. They argued against the presence of extended oceans on the basis that Titan's orbit would have been circularised by the dissipative motions of their tides. This left only one viable hypothesis: a surface where the liquid was confined to sparse, disconnected pockets. In due course, Cassini's radar was able to construct a high resolution map of the surface, by piercing the haze of Saturn's largest moon. These observations vindicated the theoretical predictions in spectacular fashion \citep{stofan2007lakes}. The liquid hydrocarbons on Titan appear to account for little more than one per cent of the total surface area.  

We are currently faced with the even more daunting task of characterising the surfaces of habitable exoplanets. But one subtlety which appears to have been overlooked is that the prediction of \citet{dermott1995tidal} could have been made \emph{even without the orbital data}. On a purely statistical basis, and in the absence of correlations, one expects the division of liquid and solid surface areas to be highly \emph{asymmetric}.   This is because the volume of liquid need not match the capacity of perturbations in the solid.  The two quantities often differ by several orders of magnitude. If it is the liquid which dominates, the solid surface becomes completely immersed.  Enceladus and Europa offer exemplary cases of this phenomenon. Beneath each of their icy crusts, a single ocean completely envelops a solid core \citep{kivelson2000galileo, waite2009liquid}. 
If, on the other hand, the liquid's volume is subdominant, it settles into small disconnected regions, as was found to be the case on the surface of Titan. 

Does this trend of asymmetric surface partitions extend to habitable exoplanets? And if so, why do we observe the Earth's water and land areas to be so finely balanced, differing in extent by only a factor of two? These are the core questions we shall aim to address in this work. 

Simulations of terrestrial planet formation provide us with the first clues for solving these puzzles. \cite{raymond2007high} explored the viability of delivering water to habitable planets from icy planetesimals which originate beyond the snow line. The chaotic nature of this process ensures habitable planets garner a broad spectrum of water compositions. This variety reinforces our expectation that their surfaces tend to be dominated by either solid or liquid.   However not all water will reside on a planet's surface. Some will remain locked in the mantle, while a further portion will be lost through the upper atmosphere.   Indeed a number of processes can influence the depths of the oceans \citep{1985Schubert, mcgovern1989thermal, kasting1992oceans, holm1996hydrothermal, 2012Abbot, cowan2014water}.   If sufficiently strong feedback mechanisms are at work, it may be possible to ensure that the depths of oceans match the amplitude of perturbations in the crust.  In which case, we ought to expect many habitable planets to resemble the Earth's division of land and sea. However it remains unclear if any are strong enough to correct for variations in water volume of more than one order of magnitude.  Alternatively, habitable planets display a broad distribution of surface conditions, and for the case of the Earth we just `got lucky'  \citep{cowan2014water}.  But, given that trillions of dice have been rolled, do we require any luck at all? Perhaps the dice were weighted in favour of a balanced surface. 
 
The earliest applications of anthropic selection were of a binary nature, in that they addressed the question of whether a particular set of conditions forbade our existence \citep{1974IAUS...63..291C, carter1983anthropic,  barrow1986anthropic, 1987PhRvL..59.2607W}. Later, more refined studies invoked Bayesian statistics to deliver quantitative assessments of how our cosmic environment may be biassed by our existence \citep{1995MNRAS.274L..73E, 2004Garriga, tegmark2006dimensionless, 2007PeacockAnthropic, simpson2016habitable, 2015SimpsonAliens}.  Until recently, the application of Bayesian anthropic reasoning was restricted to the cosmological realm.  The hypothetical ensemble of cosmic conditions has a number of theoretical motivations, yet any experimental evidence lies tantalisingly beyond our grasp. No such limitations exist for the ensemble of habitable planets.
 \citet{2015SimpsonAliens} used a simple population model to argue that our planet is likely to be towards the large end of the spectrum,\footnote{See also the pedagogical animation by MinutePhysics: \\ \url{https://youtu.be/KRGca_Ya6OM}} 
 inferring the radius $R$ of a given planet with intelligent life to be $R < 1.2 R_\oplus$  (95\% confidence bound). Empirical analyses by \citet{2015Rogers} and \citet{2016ChenKipping} appear to support these findings, with the latter study concluding that the Terran-Neptunian divide occurs at approximately $1.2 R_\oplus$. 
 Whether its the multiverse, extra-terrestrial life, or even the longevity of our species \citep{gott1993implications, 2016SimpsonDoomsday}, putting this predictive framework to the test is rarely practical.  Yet the characterisation of habitable exoplanets provides a remarkable opportunity to do just that. 
 
 %  predictive framework 
 % put this predictive framework to the test.
 % These anthropic selection effects have a number of applications including studies of cosmology
 %  The  characterisation of habitable exoplanets signifies a remarkable opportunity to test anthropic selection effects. These have a number of applications including studies of cosmology, the early universe,  the search for extra-terrestrial life \citep{2015SimpsonAliens}, and even in determining the fate of our species \citep{gott1993implications, 2016SimpsonDoomsday}..  

In this work we turn our attention to the selection effect involving a planet's ocean coverage. Our understanding of the development of life may be far from complete, but it is not so dire that we cannot \emph{drastically} improve on the implicit approximation that all habitable planets have an equal chance of hosting intelligent life.   Should we consider  planets with different land-ocean divides to have an equal chance of  producing an intelligent species such as \emph{Homo Sapiens}? Few would doubt whether the Earth's surface configuration is better suited to supporting a diverse biosphere than  \emph{Tatooine}.  It is this small piece of knowledge that can be exploited to update our prior belief for the surface conditions among the ensemble of habitable planets. 

In \S \ref{sec:tuning} we explore the fine-tuning problem associated with the Earth's oceans, and review two approaches for tackling the problem:  feedback processes and  observational selection effects.
In \S \ref{sec:fecund} we quantify the relative probability of observing a host planet based on its habitable area.
The model we use for the ensemble of surface conditions is defined in \S \ref{sec:model}.
Our main results are presented in \S \ref{sec:results}, before concluding with a discussion in \S \ref{sec:conclusions}.

\section{The Oceanic Fine-Tuning Problem} \label{sec:tuning}

\subsection{Basin Saturation}

To facilitate a comparison of surface conditions across the gamut of habitable worlds, it is instructive to introduce the \emph{basin saturation} $\Vnorm$. This dimensionless quantity is defined as the ratio of the  surface water volume $\Vw$ to the capacity of the basin $\Vb$,
\[ \label{eq:X}   
\Vnorm \equiv       \frac{\Vw}{\Vb}   \, . % \frac{\text{Ocean Volume}}{\text{Median Basin Capacity}}  
\]
Here we shall take the basin capacity $\Vb$ to be the volume of liquid required to cover half of the solid surface.\footnote{Note that we could equally  have defined $\Vb$ in terms of the volume required to cover 100\% of the area. However this quantity is more challenging to model as it is highly sensitive to the extreme tail of the elevation profile.} By construction,  a sub-saturated world $(\Vnorm<1)$ will have a surface dominated by land, while a super-saturated world  $(\Vnorm > 1)$ will be mostly covered by water. The Earth's saturation value is   $\Vnorm_\oplus \simeq 4$. One could generalise this expression beyond water, to encompass any liquid. For example, Titan's lakes have an estimated volume of around $9,000\,\mathrm{km}^3$, corresponding to a basin saturation value of approximately $10^{-4}$.

\begin{table*}
\caption{Characteristics of known solid surfaces.}
 \label{tab:elevations}
\begin{tabular}{ccccc} 
 \hline
     &   Radius              	&   RMS Elevation        &  Basin Capacity                                       & Basin Saturation \\ 
           & $R (\mathrm{km})$      & $\sigma_h$(km)       & $\Vb$$( 10^6 \, \mathrm{km}^3)$  &  $\Vnorm$  \\
   \hline
Moon      & 1,737 & 1.95& 59 & - \\ 
Mercury  &2,440&  1.09 & 65 & -\\ 
Titan      & 2,575 & 0.13 & 8 &  $10^{-4}$  \\ 
Mars     & 3,390             &  2.98 &   343 & -  \\ 
Venus     &  6,052  & 0.68 & 248 & - \\ 
Earth      &  6,371 & 2.51 & 1,021 & 4 \\ 
\hline
\end{tabular} 
\end{table*}

Throughout this work we shall use the term habitable planet to refer to those worlds which possess a permanent body of surface water, such that $\Vnorm>0$.   The full ensemble of habitable planets will span a distribution that we shall denote $p(\Vnorm)$.   Unless  $p(\Vnorm)$ has  \emph{both} a mean close to unity $(\mu_\Vnorm \sim 1)$ and a small standard deviation  $(\sigma_{\Vnorm} \lesssim 1)$ then most planets will have imbalanced surfaces, dominated by either  land or ocean.  The oceanic fine-tuning problem may therefore be stated as follows: Why should we find ourselves on a planet with a saturation value  $\Vnorm$  of the order unity, as opposed to $\Vnorm \gg 1$ or $\Vnorm \ll 1$? 

There are three viable hypotheses:

\begin{enumerate} 
\item $\mathcal{H}_0$: \emph{Luck.} The distribution of saturation values  $p(\Vnorm)$ is not localised at  $\Vnorm \sim 1$, yet  by chance we arrived at the point $\Searth \sim 1$. 
\item $\mathcal{H}_1$: \emph{Selection.} The distribution of saturation values  $p(\Vnorm)$ is not localised at  $\Vnorm \sim 1$, but land-based observers such as humans inhabit an inherently biased sample of habitable planets, such that the conditional distribution $p(\Vnorm | H)$ is localised close to $\Vnorm \sim 1$. This scenario is explored in Section \ref{sec:select}.
\item $\mathcal{H}_2$: \emph{Feedback.} The distribution $p(\Vnorm)$ is localised close to $\Vnorm \sim 1$ due to feedback mechanisms.  This hypothesis is discussed in Section \ref{sec:feedback}.
\end{enumerate}

% Why, then, does the Earth have a balanced surface? 
% why should we find ourselves on a planet with a saturation value  $\Vnorm$  of the order unity, as opposed to $\Vnorm \gg 1$ or $\Vnorm \ll 1$?    Should this value of $\Vnorm_\oplus = 4$ be a cause for concern?  To see why it ought to be, 
% susceptible
Why might the terrestrial value, $\Searth$, be prone to selection bias? Consider the relationship between the basin saturation value $\Vnorm$ and the fractional oceanic area $\oceanfrac$. This relationship is illustrated by the solid line in the left hand panel of Figure \ref{fig:landfrac}, for the case of a Gaussian elevation profile. The dot-dashed and dashed lines utilise the elevation profiles of the Earth and Mars respectively.  Meanwhile the shaded region $(\Vnorm > 10)$  denotes the regime in which over $99.7\%$ $(3 \sigma)$ of a Gaussian surface is immersed. This fraction will change slightly for different elevation profiles, but only the most contrived shapes would retain a substantial land mass. (For example, if we tried to carve out more room for the Earth's oceans, by excavating 2/3rds of the continental land mass and replacing it with water, this would raise our basin saturation value from 4 to 6). The Earth's saturation value, as represented by a vertical dashed line, sits close to the threshold at which planets transition to water worlds.  
Is it just a coincidence that we are located close to a critical point, beyond which our existence would not have been possible?
Coincidences often arouse our suspicion, but this is an intuitive response, one that is difficult to quantify. Fortunately Bayesian statistics offer a means to analyse  and quantify the source of this distrust \citep[see e.g.][]{mackay2003information}. 

% ``Why do coincidences make us suspicious?" - D Mackay

Note that for any Gaussian elevation profile, the expression (\ref{eq:X}) for the basin saturation has the generic solution
%\[ \label{eq:gaussX}
%\eqalign{
%\Vnorm &= \frac{ \int_{-\infty}^z \erfc(x) \ud x  }{   \int_{-\infty}^0 \erfc(x) \ud x }\, , \cr
%&=     \exp \left[ - \frac{z^2}{\sqrt{\pi}} \right]  - z \sqrt{\pi}     \erfc(z)  \, ,
%}
%\] % , the terms of standard deviations
\[ \label{eq:gaussX} % 
\Vnorm = \frac{ \int_{-\infty}^z \frac{1}{2} +  \frac{1}{2} \erf(x)   \ud x  }{   \int_{-\infty}^{0} \frac{1}{2} +  \frac{1}{2} \erf(x) \ud x }\, =  \mathrm{e}^{-z^2}   +  z \sqrt{\pi}     \erf(z)  \, ,
\] % , the terms of standard deviations
where $z  \equiv   \sqrt{2} \erf^{-1}\!{\left(1 - 2 \oceanfrac \right)} $. Here $\erf$ denotes the error function, while the parameter $z$ represents the sea level's displacement from the median elevation.

% Note that by imposing this age restriction we are implicitly excluding all planets around high mass stars $M \gtrsim 1.4 M_\odot$. 
Both $\Vw$ and  $\Vb$, and therefore $\Vnorm$, will exhibit some time-dependence over geological timescales. To draw a fair comparison across different planets, we must therefore specify a fixed reference point. Here we shall concern ourselves with their values at an age of $4\,$Gyr, the approximate time required for the emergence of land-based and intelligent life on Earth. Therefore Mars, for example, would be considered to have $\Vnorm = 0$, despite its possible early period of habitability. Note that by imposing this age restriction we effectively exclude planets hosted by higher mass stars,  $M \gtrsim 1.4 M_\odot$. By this time these stars will have evolved off the main sequence, posing a serious challenge for habitability \citep{2016ApJ...823....6R}. Planets within lower mass M-dwarf systems are included, unless the emergence of complex life has been compromised by their heightened stellar activity.

With this definition, we expect to find a bimodal distribution for $\Vnorm$. Those planets which lose water on a timescale much less than $4\,$Gyr will be deemed uninhabitable, $\Vnorm = 0$, while those capable of retaining water shall form a broader distribution $p(\Vnorm)$.

\subsection{The luck of natural selection} \label{sec:select}

% Aside from feedback mechanisms, what reason might there be behind our precariously laden oceans? 
As is evident from Figure \ref{fig:landfrac}, the Earth appears precariously close to the waterworld limit. This marks the transition to a regime where the existence of our species would no longer be viable.  Such proximity to a critical limit is exactly what one expects to find, under one condition: the bulk of the probability distribution lies beyond the critical point. In other words, if we cannot exist on a waterworld, yet most habitable planets are waterworlds, then we should expect to live on a planet close to the waterworld limit. This is the same line of reasoning used by  \citet{1987PhRvL..59.2607W} to predict the value of the cosmological constant. 

Given how closely the cosmic argument parallels our planetary one, it is worth revisiting the logical steps followed by \citet{1987PhRvL..59.2607W}. There may be an ensemble of cosmic conditions, and this ensemble defines a probability distribution for the cosmological constant, $p(\Lambda)$.  Values of the cosmological constant greater than some critical value $\Lambda_c \sim 10^{-120}$ prohibit the formation of galaxies. If most values of the cosmological constant are too large to permit life, then the selection effect associated with our existence will truncate most of the probability distribution $p(\Lambda)$, such that $p(\Lambda>\Lambda_c) = 0$.  Despite the very large uncertainty in the functional form of $p(\Lambda)$,  a single sample ought to lie close to the point of truncation, provided the tail of the distribution is smooth and featureless.  It was this statistical insight that led Weinberg to conclude that the value of the cosmological constant in our Universe is within an order of magnitude of the critical value required to obstruct galaxy formation. Empirical verification arrived little more than a decade later  \citep{1998Riess, 1999Perlmutter}.

% What  is that 
Returning to the case of planetary oceans, we are faced with a somewhat analogous situation. In place of  $\Lambda$, we now consider the influence of a planet's ocean coverage. Given that our existence would not have been tenable on waterwolds, this imposes an upper bound given by $\Vnorm_c \simeq 10$. In which case, the selection effect truncates the full $p(\Vnorm) $ distribution, such that $p(\Vnorm>\Vnorm_c) = 0$. If the bulk of habitable planets lie beyond the threshold - i.e. they are waterworlds -  then we should fully expect to find that our home planet lies in the range $1 < \Searth < 10$.   Conversely if the bulk of habitable planets fall below the threshold, such that waterworlds are outnumbered, then we have no immediate expectation that our planet should lie in close proximity to the threshold. 

There is an important difference between the cosmological and planetary inferences.  For the case of the cosmological constant, the hypothetical ensemble was used to predict the local value.  For the case of planetary oceans, the information passes in the opposite direction: it is our local value which is being used to predict the nature of the ensemble. That is the core concept which underlies this work.

% To proceed in a quantitative fashion, % 1995MNRAS.274L..73E, tegmark2006dimensionless
% This 'black-or-white' approach to habitability  is useful, but we can go further. 

For a further example on the importance of selection effects, we turn to biology.  Early civilisations assumed a creator was responsible for all of the highly complex designs exhibited by living creatures. These designs, it turned out, could be explained by a mechanism of natural selection \citep{matthew1831naval, darwin1872origin}. This went on to become one of the most famous and widely accepted results in science. 

Evolution determined which genes we call our own, but what determined which planet we call home? If one wishes to avoid invoking a creator, then one must accept that a higher tier of natural selection took place  -  on a truly cosmic scale.  Unlike the animal kingdom, where genetic material undergoes sequential generations, planetary selection is a shotgun approach. Only a single `generation'  exists. But the planetary population is vast, with their broad ensemble of characteristics mimicking the range of genetic mutations. The end result is highly analogous. Our genes, and our planet, are those which have proven to be highly successful at producing life. In biology, we cannot see those genetic mutations which are associated with sterility. In the same way, no individual in the Universe evolved on a planet whose characteristics are associated with sterility. From this deep selection process grows the appearance of design.  

The apparent fine-tuning of the Earth's orbit - that it is neither too close to the Sun for its oceans to boil, nor too remote for them to freeze - is readily attributed to the importance of liquid water in the development and sustenance of life. Could the Earth's ocean coverage be a further example of illusory design? The land-ocean divide is likely to have a major impact on the probability of forming an intelligent land-based species such as our own.  Planets with only small areas of exposed land  will have a much more limited range of land-based species, and this prospect vanishes altogether in the case of total ocean domination.  Conversely, consider a planet identical to the Earth, except it has only sufficient surface water to fill the Mariana Trench. It is still technically classified as habitable, but would it be as likely as the Earth to produce a species such as ourselves? There would only be a tiny area of habitable land, while the remainder is arid desert.  

Establishing a selection process based on land-based species does \emph{not} discount the plausibility of water-based observers. There may be a number of water-based and land based observers, but \emph{a priori} it is extremely unlikely that these two numbers are a similar order of magnitude. And since we find ourselves to be land-based observers, it is highly probable that we are vastly more numerous than any water-based counterparts.

\subsection{Feedback Mechanisms} \label{sec:feedback}
If most habitable planets possess an approximately even divide between land and oceans, despite a variety of initial conditions at the time of their formation, then some process or combination of processes must act to equilibrate the ocean and basin volumes.  Here we briefly review some of the processes which are capable of relating these quantities. 

\emph{Isostatic Equilibrium:}
To a certain extent, the oceans make room for themselves by exploiting their own weight. Deeper, heavier oceans impart a higher pressure on the seabed, which pushes the crust lower into the mantle. The magnitude of this effect is proportional to the water-to-mantle density ratio, which in the case of the Earth is approximately one third. So if the Earth's oceans were to suddenly vanish, seabeds would rise by an average of one kilometre. The influence of this feedback mechanism is illustrated in the left hand panel of Figure \ref{fig:landfrac}. The solid line shows how the fractional ocean coverage would evolve for different quantities of surface water, when fixing the Earth's elevation profile. The dotted line shows the corrected curve, taking into account the effect of isostasy. The two lines are only distinguishable within a narrow regime, where the oceans are a comparable depth to the continents. Isostacy therefore cannot help make substantial corrections to the land-ocean divide.

\emph{Deep Water Cycle:}
Water is recycled between the oceans and the crust: it is emitted at mid-ocean ridges, and returned via the subduction of tectonic plates. What is less well understood is the extent to which water is transported deeper into the mantle.  \cite{kasting1992oceans} proposed that an exchange of water between the crust and mantle could act as a buffer, preventing the oceans from becoming much shallower than their current depths \citep[see also][]{hirschmann2006water}.  \citet{cowan2014water} present a model which accounts for the stronger surface gravity on super earths, which suggests larger terrestrial planets could maintain an exposure of land.
While these processes certainly have the potential to provide significant feedback effects, the capacity of the Earth's mantle is thought be within a factor of ten of the oceans \citep{inoue2010water}. It therefore cannot correct for order-of-magnitude fluctuations in water composition. % Meanwhile the total capacity of the mantle  \citep{inoue2010water}. 

%  While these processes certainly have the potential to provide significant feedback effects, the inventory of the Earth's mantle is thought be within a factor of a few of the oceans. It therefore was not responsible in correcting for order-of-magnitude fluctuations in the Earth's water composition. % Meanwhile the total capacity of the mantle  \citep{inoue2010water}. 

\emph{Self Arrest:} If large excesses of water aren't stored in the mantle, the alternative disposal route is via the upper atmosphere.  \citet{2012Abbot} propose that some water-dominated planets may initiate a `moist greenhouse' phase, which endures until they have lost sufficient water to expose continents.  At this point, CO2 is sequestered via silicate weathering, thereby bringing the climate under control. This model has the appealing property that an approximately even land-ocean divide will serve to maximise the weathering rate, as a balance of exposed land and precipitation is required.   The climactic impact of a high CO2 concentration has been explored by \citet{kasting1986climatic} and more recently \citet{2014Ramirez}, who demonstrate that the Earth's atmosphere is likely to be stable against a runaway greenhouse effect, potentially allowing for a controlled release of excess water. However a consensus on the matter has yet to be reached. \citet{wordsworth2013water} argue that atmospheric cooling effects may act to limit the escape of $\text{H}_2\text{O}$ in most cases. They conclude that significant water loss through the upper atmosphere may only occur under special conditions.
% 

%  In any case, even if self-arrest does occur, it still remains unclear why  $5 \times 10^8  \, \text{km}^2$ is the critical area of exposed land required to do so, and not  $5 \times 10^7 \, \text{km}^2$ or $5 \times 10^9 \, \text{km}^2$.
% Whether the Earth experienced such a phase in its early history remains an open question. 

\emph{Erosion \& Deposition:}
One of the more distinctive features seen in the left panel of Figure \ref{fig:landfrac} is the sharp drop in land area associated with a modest rise in the current sea level. This feature has been associated with the processes of erosion and deposition \citep{rowley2013sea}. This reduces the amplitude of perturbations, and leads to a build up of material close to sea level. The extent to which it can influence area coverage is unclear. As with isostacy, its efficacy is limited to the regime where a balance between land and ocean areas is already in place. 

In summary, while each of the above mechanisms contribute to the complex relationship between elevation and the volume of surface water, it remains unclear that they are strong enough to equilibrate the diverse conditions of habitable planets. Aside from variability in their water composition, planets will also display variable crust compositions, a range of surface gravities, and differing degrees of tectonic activity, all of which will contribute to a broad variety of elevation profiles. That being the case, in this work we shall focus on the prospect that habitable planets display a broad range of surface water-to-basin volume ratios, such that $\sigma_\Vnorm/\Vnorm$ is of the order unity. 

\begin{figure}
\includegraphics[width=78mm]{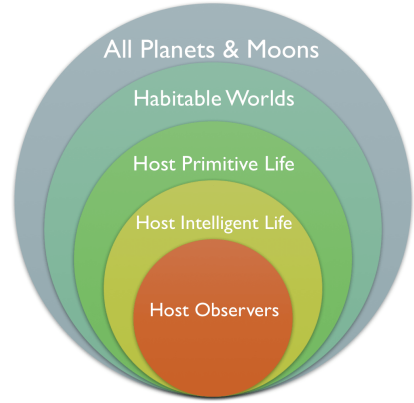}
\caption{A schematic diagram of the nested hierarchy of planets based on their biological status. Progressively deeper subsets are associated with environments which are increasingly well suited to nurturing the development of living organisms.}
\label{fig:sets}
\end{figure}

\section{Terrestrial Selection Effects} \label{sec:fecund}
%We wish to interpret one of the Earth's characteristics, $\thetaearth$, within the context of the full ensemble of planets in the Universe, $p(\theta)$.

% How can we quantify the observational selection effect on a given planetary characteristic $\theta$? 
How can the terrestrial value of some parameter,   $\thetaearth$, be used to inform us of the full ensemble of planets in the Universe, $p(\theta)$? For an unbiased sample, then $\thetaearth$ represents an unbiased estimator of the population mean. However the stringent conditions required for intelligent life to evolve on a planet are likely to impose a bias on $\thetaearth$. In order to quantify this effect, we turn to  Bayes' Theorem
\[ \label{eq:bayes}
p_\oplus(\theta) \propto p(\theta) p(\observer | \theta) \, ,
\]
Here  $p(\theta)$ denotes the probability distribution of $\theta$ across all planets,  while   $p_\oplus(\theta)  \equiv  p(\theta | \observer)$ denotes the probability of an observer's host planet having a parameter $\theta$. Meanwhile  $p(\observer | \theta)$ is the term responsible for inducing a selection effect. It is often neglected or forgotten because it can be challenging to estimate, yet it is categorically incorrect to do so \citep{tegmark2006dimensionless}.   Even the most innocuous of parameters, such as the proportion of Argon in the atmosphere, are not immune from this selection bias. If it is correlated  with any variable that influences the formation or development of life, such as the abundance of water or hydrocarbons, then a selection bias will arise. Therefore, in principle at least, the observed value of almost any of the Earth's features could lie in the far extremities of the full distribution of habitable planets.

\subsection{Planetary Fecundity} 
 
 A planet's fractional ocean coverage $\oceanfrac$ has a major influence on the area available for land-based species to evolve and  thrive, and therefore it is likely to play a significant role in the emergence of intelligent species.  In this section we shall therefore aim to model the selection effect associated with a planet's habitable land area $H$.  This necessitates a statistical description for the  evolution of intelligent life. This may seem like a hopeless endeavour, due to the vast uncertainty in the amplitude of the probabilities involved. However, we are only interested in the selection bias, so the overall normalisation is irrelevant. Selection effects are only sensitive to relative changes, not how rare or abundant life is on the whole. 
   
\begin{figure*}
\includegraphics[width=175mm]{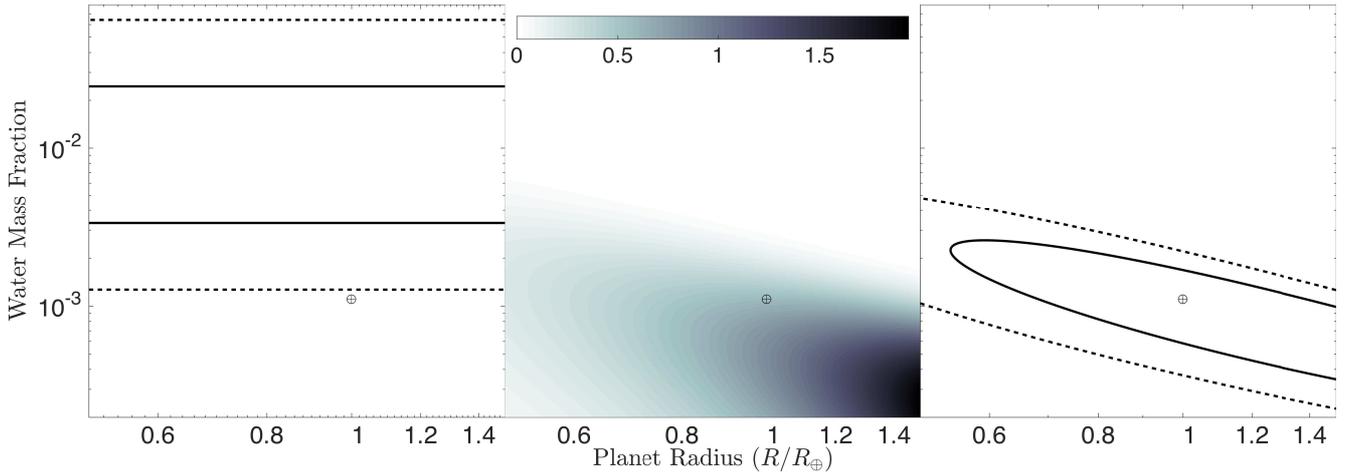}  
\caption{An illustration of the strong bias which can arise for the radius $R$ and water mass fraction $\WMF$ of an observer's host planet. The Earth is represented by the black square. \emph{Left:} Our model for the joint probability distribution $p(R, \WMF)$, where the size distribution of habitable planets $p(R)$ is uniform in log space while water mass fraction is Gaussian in log space. (68 and 95 per cent CL). \emph{Centre:} The habitable land area $H$ available on a given planet, in units of the Earth's land area (approximately 149 million $\mathrm{km}^2$).  Larger planets are particularly prone to becoming ocean dominated. \emph{Right:} The joint probability distribution for an observer's host planet, $p(R , \WMF | \observer)$, derived from the quantities depicted in the other two panels, via equations (\ref{eq:bayes}) and (\ref{eq:habitability}). (68 and 95 per cent CL).}   
\label{fig:bias}
\end{figure*}

Figure \ref{fig:sets} depicts the nested sequence of selection processes that must be disentangled if we are to identify those planets which are likely to harbour life. Each set holds its own biased distribution of parameters $p(\theta)$, and these distinguishing features are helpful in the search for extra-terrestrial life. The relationship between the two outermost sets defines the study of habitability. In other words, what characteristics must a world possess for it to be habitable?  In this work we shall focus on the link between the innermost set (observers) to the middle set (primitive life). This defines a quantity we shall refer to as the fecundity. It tells us what life-bearing worlds require in order to produce sentient individuals. This relationship also informs us how we should expect our planet to differ from the bulk of life-bearing worlds. This can be formalised using a variant of equation (\ref{eq:bayes})  
 
 %Initially, observations of exoplanets are drawn from the full outer ensemble, with their own set of observational biases, yet we wish to focus our attention on the smaller subsets. Each subset, which may comprise a very small fraction of the total, holds its own biased distribution of parameters $p(\theta)$, and these distinguishing features are helpful in the search for extra-terrestrial life.We only have one sample within the life-bearing set. However it has been sampled from the innermost category, and is therefore unlikely to be representative of the planets which harbour primitive life.  Nonetheless, even just a qualitative understanding of the selection effects is sufficient to provide us with insight into the nature of other life-bearing worlds.  
 
  \[ \label{eq:bayeshabitable}
 p_\oplus(\theta) \propto  p_L(\theta) F(\theta)   \, ,
 \]
where $p_L(\theta)$  is the distribution of $\theta$ among all life-bearing planets, and $F(\theta)  \equiv  p(\observer | \theta, L) $ is the fecundity.  To emphasise the important distinction between fecundity and habitability, consider the case of Enceladus. Its sub-surface oceans are widely considered to be a promising place to search for microbial life. But no-one is under any pretence that intelligent life forms could be thriving in such a confined and energy-scarce environment. It may be habitable, but it is lacking in fecundity.   

In order to estimate a planet's fecundity, it is helpful to decompose it into two separate components
\[  \label{eq:inhabitable}
F(\theta)  =   f_i(\theta)  \bar{N}_i(\theta)  \, ,
\]
where  $f_i(\theta)$ is the proportion of life-bearing worlds upon which intelligent life evolves, and  $\bar{N}_i(\theta)$ is the mean number of observers produced by life-bearing planets with parameter $\theta$. Note that the normalisation of $F(\theta)$ is irrelevant, only the manner in which it evolves with $\theta$ will influence (\ref{eq:bayeshabitable}).
The  formalism above is valid  when adopting either the self-sampling assumption \citep{bostrom2002self} or the self-indication assumption \citep{bostrom2003doomsday}, because we are working within a local ensemble.

\subsection{Fecundity of the Habitable Land Area}

In principle this formalism can be applied to any planetary parameter. In this work we shall focus on the habitable land area, $H$.  We model the emergence of an intelligent species within a given area of habitable land as a rare stochastic event. Larger areas of habitable land permit a greater abundance and diversity of organisms to explore the evolutionary landscape. There is therefore a greater opportunity for one species to undergo a period of prolonged encephalisation, and ultimately form an intelligent species.  This model suggests that the evolution factor from equation (\ref{eq:inhabitable}) exhibits a linear scaling of the form $ f_i(H) \propto H$.

Once an intelligent species has become established, we assume it spreads to occupy the available habitable land area. Therefore the mean number of individuals is also likely to scale in proportion with the habitable land area, yielding $N_i(H) \propto H$.   The reason for its inclusion here is that any given individual (such as yourself) is more likely to reside on a more populous planet. This may be an unsettling statement, but it is no different to stating that you are more likely: to have a common blood type compared to a rare one; to live in a high population country than a small one; to travel on a busy train than a quiet one. These are all intuitive concepts, and (en masse) they represent experimentally verifiable statements. They are not based on human behaviour but simply the variance of group sizes, coupled with our personal status as an ordinary individual. In general, any group of which you are a member does not provide an unbiased estimate of the median group size - it is an overestimate. There is little reason to believe this trend should or could stop abruptly at the planetary scale. 

To summarise, we model the two factors from (\ref{eq:inhabitable}) as  $f_i(H)   \propto H$ and $N_i(H) \propto H$, 
which combine to yield a quadratic scaling relation for the fecundity
\[ \label{eq:habitability}
F(H)  \propto H^2 \, .
\]
For our fiducial model, the fecundity of a planet is taken to be proportional to the square of the habitable land area. One factor of $H$ originates from the higher probability of successfully forming an intelligent species somewhere on the surface, if there is greater available area. The second factor stems from the higher mean population size associated with that greater area.  We have also verified that our findings are not significantly modified if we replace our fiducial $F(H)  \propto  H^2$ model with linear or cubic scaling relations. Therefore the results presented in this work hold even if the formation of intelligent life is an easy process \citep{simpson2016habitable}.

\subsection{Modelling the Habitable Land Area}
 
We can express a planet's habitable land area as follows
\[ \label{eq:H} 
H  = 4 \pi R^2 (1 - \oceanfrac)  \habfrac \, , 
\]
where $\habfrac$ is the fraction of land which is deemed to be habitable.   The habitable fraction must vanish, $\habfrac \rightarrow 0$, as we approach the limit where there is no surface water $(\oceanfrac \rightarrow 0)$. In this extreme case the entire surface is rendered an uninhabitable desert. Planets with progressively greater ocean coverage will, on average, possess a diminishing proportion of desert. Ideally a suite of climate simulations could be used to estimate the shape of this function. In this work we interpolate these two extremes using a power law
\[ \label{eq:desert}
\habfrac  =  {\oceanfrac}^\waterconst  \, ,
\]
where the Earth's values of $\habfrac \simeq 0.7$ and $\oceanfrac \simeq 0.71$ are consistent with our fiducial value $\waterconst = 1$. 
In the right hand panel of Figure \ref{fig:landfrac},  the three lines correspond to the habitable land areas for three different $\waterconst$ values:  $\frac{1}{5}, 1$, and $5$.

The oceanic coverage $\oceanfrac$ is computed via the hypsometric curve $\oceanfrac = \hyps(h_w)$, where the sea level $h_w$ is found by integration
\[ \label{eq:Vw}
\Vw = 4 \pi R^2 \int_{-\infty}^{h_w}  \hyps(h) \ud h \, ,
\] 
provided $|h| \ll R$. 

 \section{The Ensemble of Surface Conditions} \label{sec:model}

In this section we construct a model for the ensemble of oceanic and land areas among habitable planets. This model will be used in the following section to illustrate the selection effect that we are susceptible to when we measure our host planet's ocean coverage.

Since we have already established a relationship between the saturation value $\Vnorm$ and the oceanic area, we seek to identify the two components that define $S$, namely the oceanic volume $\Vw$ and the basin capacity $\Vb$. 

 \subsection{Volume of the Oceans}
 
 The volume of surface water on a given planet may be decomposed into four contributing factors 
 
 \[
 \Vw = \WMF M  f_s \rho_w^{-1}  \, ,
 \]
 where $\WMF$ is the water mass fraction, $M$ is the mass of the planet, $\rho_w$ is the density of water, and $f_s$ is the fraction of the planet's water inventory which resides on the surface.  In principle each of these variables could evolve as functions of time. For simplicity, and to permit a consistent comparison between different planets, we take these values to refer to the planetary composition at an age of $4\,$Gyr. That different planets will lose water at different rates will likely serve to increase the variance in their water compositions, compared to their initial states. 
 
To estimate the distribution of water mass fractions $p(\WMF)$ across the ensemble of habitable planets, we use the 15 planets from the numerical simulations of \citet{raymond2007high}, as presented in their Table 2. We find no statistical evidence of a correlation between the size of a planet and its water mass fraction. To determine whether the data vector is consistent with being sampled from a Gaussian, we employ the Anderson-Darling test. The raw $\WMF$ values were found to be incompatible with a Gaussian,  while $\log(\WMF)$ was found to be consistent.  We therefore model $p(\log \WMF)$ as a Gaussian distribution, with a mean and standard deviation motivated by the 15 planets in the simulation: $  \log(9\times 10^{-3})$ and $ 0.8$ respectively.  Estimates for the Earth's  value of $\WMF$ vary considerably. Throughout this work the terrestrial value is taken to be $\WMF_\oplus = 10^{-3}$ (but note that this  fiducial model is only adopted for illustrative purposes, it has no bearing on our final results).
 
 To estimate $M$ we adopt an empirically determined mass-radius relation $R \propto M^{0.28}$ \citep{2016ChenKipping}.  Based on projections from Kepler data \citep{2015Silburt}, we consider  the planetary  radii to be  evenly distributed in log space, $p(R) \propto R^{-1}$.  The breadth of the distribution $p(R)$ is not of particular significance to this work. Here we adopt a relatively broad range of $0.5 < R/ R_\oplus<1.5 $ in order to illustrate possible radius-dependent effects.  
 
We consider the surface water fraction $f_s$ to be constant in our fiducial model.  Stochastic fluctuations in $f_s$ are entirely degenerate with fluctuations in $W$, so can be absorbed into $\sigma_w$. If $f_s$ were to change systematically as a function of the planet's surface gravity, as explored in the Appendix, this does not appear to have a significant impact on our results.  However we would expect significant changes if $f_s$ were correlated with the water mass fraction $W$,  One example of this correlation is where the mantle reaches a saturation point, meaning that values of $W$ beyond a critical point lead to ever-increasing values of $f_s$. Meanwhile, at some critically low value of $W$ one expects there to be no permanent surface water at all. For a given planet, some minimum volume of water is required to saturate the surface environment. It is unclear where this critical `desertification' point lies, here we shall simply assume this value lies below the range of values under consideration.  If at moderate values, between these extreme regimes, an increase in $W$ leads to a decrease in $f_s$,  this would be indicative of a regulatory feedback mechanism, categorised earlier as hypothesis $\mathcal{H}_2$. As discussed earlier, this could only operate over a modest range, due to the limited capacity of the mantle. Here we shall focus on exploring the viability of hypothesis $\mathcal{H}_1$.

\begin{figure*} 
\includegraphics[width=80mm]{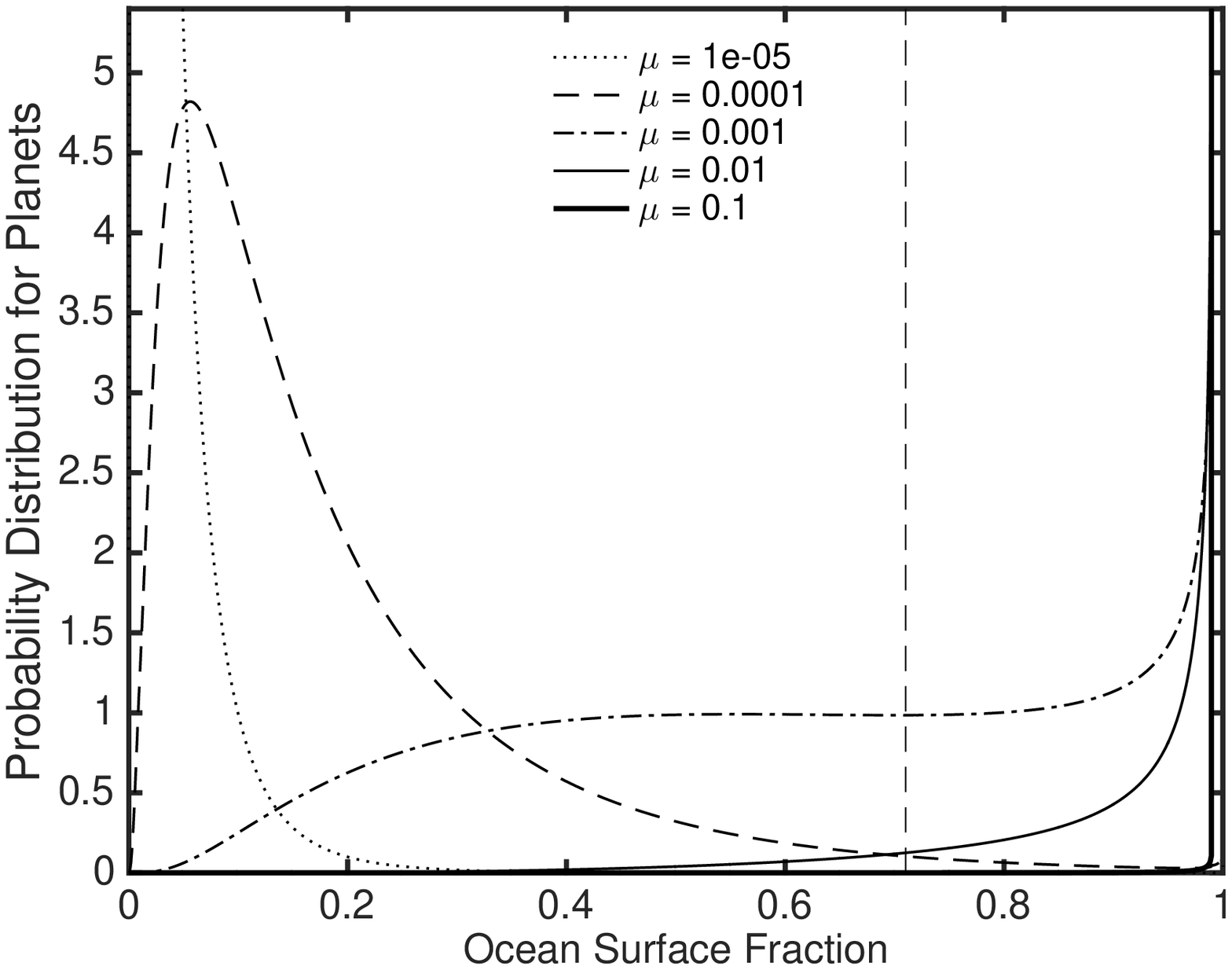} 
\hspace{0.2cm}
\includegraphics[width=80mm]{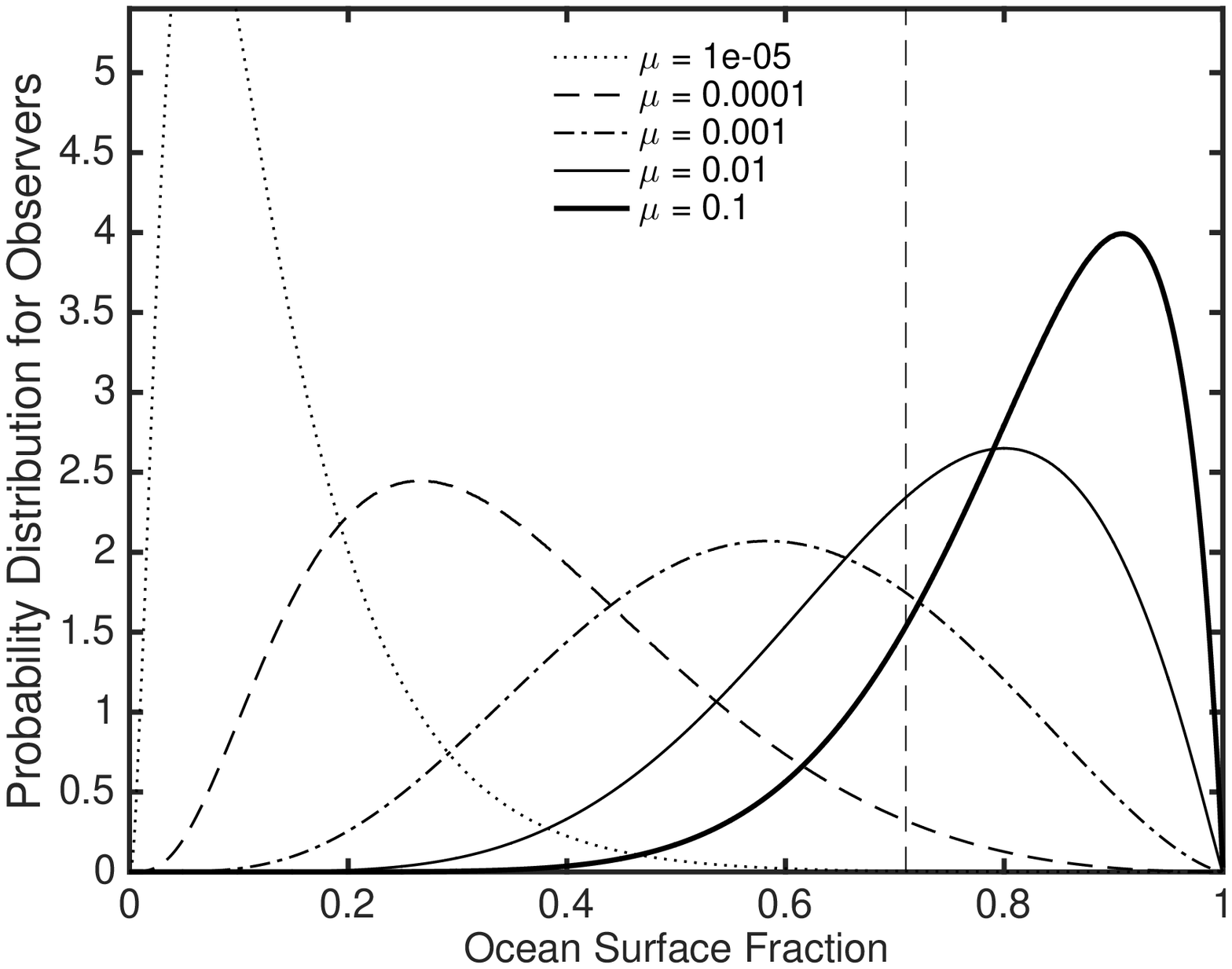} 
\caption{The influence of selection effects in determining the  ocean coverage of host planets of land-based species. 
\emph{Left:} 
The  probability distribution for the  fractional ocean coverage among habitable planets, for five different values of the median water mass fraction $\mu$.  If taken at face value, only values very close to $\mu = 10^{-3}$ are consistent with the Earth's ocean fraction of around $71\%$, as denoted by the vertical dashed line. The variance of $p(\log \WMF)$ remains unchanged in each case.
\emph{Right:}  The same five models as the left panel, but here we illustrate the probability distribution for an observer's host planet. A much broader range of models are now consistent with the observed value, particularly those in which the Earth is a relatively dry planet.
}
\label{fig:obsoceanpdf}
 \end{figure*}
 
 \subsection{Volume of the Surface Perturbations}

The hypsometric curve $\hyps(h)$ encapsulates the proportion of a planet's surface area which lies below a given elevation $h$.  While we expect planet-to-planet variability in both the shape and amplitude of the curve - the Earth shows signs of a bimodal distribution - what is of importance in this context is simply the statistical average.    We model the shape of the hypsometric curve using the cumulative distribution function, corresponding to a Gaussian elevation profile, such that
 \[ \label{eq:erf}
\hyps(h) = \frac{1}{2} +  \frac{1}{2} \mathrm{erf}{\left(\frac{h}{\sqrt{2} \stdElevation} \right)} \, .
\]
Substituting (\ref{eq:erf}) into (\ref{eq:Vw}), while setting  $h_w=0$, yields a basin capacity
%\[
%\Vb  =  4\sqrt{2 \pi}  R^2 \sigma_h \, .
%\]
\[
\Vb  =  \sqrt{\frac{2}{\pi}} 4 \pi   R^2 \sigma_h \, .
\]
Our fiducial value for the standard deviation of perturbations,  $\sigma_h$, is taken to match that of the Earth  $(\stdElevation = 2.51$km). Note that the choice of fiducial model is only for the purposes of illustration: it will have no bearing on our final results.  

Among the solar system's terrestrial planets, there is no clear trend to suggest how the amplitude of elevation profiles change with respect to the planet's radius. Even the Moon possesses deviations in elevation which are of similar magnitude to those of the Earth (see Table \ref{tab:elevations}).  For large radii, $R>R_\oplus$, the amplitude of $\hyps(h)$ is expected to decay, partly due to the stronger surface gravity, prohibiting large perturbations in the crust  \citep{2009Kite}. Yet even if no change occurs, as we shall conservatively assume in our model, planets with larger radii will experience progressively greater ocean coverage. 

 \subsection{Saturation Value} \label{sec:Vnorm}
 
Now we can combine these two expressions for $\Vw$ and $\Vb$, in order to gain insight into how the basin saturation $\Vnorm$   varies across the ensemble. 
 
 \[  \label{eq:saturation}
 \Vnorm = \frac{\WMF M  f_s \rho_w^{-1} }{4\sqrt{2 \pi}  R^2 \sigma_h } \propto \frac{\WMF M^{0.44} }{ \sigma_h } 
 \]
This allows us to estimate the standard deviation $\sigma_\Vnorm$,  a very important quantity as it dictates the magnitude of the selection bias. If $\sigma_\Vnorm$ is very small then all habitable planets would have the same surface conditions, so no selection effect can push us far from the median value.  It may be estimated by summing the contributing terms in quadrature, as follows
\[ \label{eq:sigma}
\left( \frac{\sigma_{\Vnorm}}{\Vnorm} \right)^2  \simeq  \left( \frac{\sigma_w}{W} \right)^2 +  \left( \frac{\sigma_\sigma}{\sigma_h} \right)^2 + 0.19 \left( \frac{\sigma_m}{M} \right)^2  \, ,
\]
provided the covariances are subdominant. Below we briefly discuss these three contributing factors in turn

%
%where $\sigma_{\mathrm{ws}}$ denotes the covariance between $\Vw$ and $\Vb$.
% water volumes $\sigma_w$ is sourced by many factors: variation in planetary mass; variation in
Fluctuations in the water mass fraction, $\sigma_w$, can arise via  the different compositions among proto-planetary disks, and the stochastic nature of water delivery. The simulations only account for the latter, and as mentioned earlier, they appear to have a standard deviation in log space of $0.8$, which translates to $\sigma_w/W \simeq 0.95$. 

To estimate the variability in $\sigma_h$, denoted $\sigma_\sigma$, we turn to the rocky bodies in the solar system. In Table \ref{tab:elevations}  we present the amplitude of various elevation profiles, as given by \citet{lorenz2011hypsometry} and  \citet{2016LPI....47.2959B}.  The scatter is suggestive of a fractional range $\sigma_\sigma / \sigma_h \simeq 0.8$.

Finally, for our array of habitable masses,  $\sigma_m / M \simeq 0.6$, but this does not make a   significant contribution to $\sigma_{\Vnorm }$ as it is suppressed by a factor of five. Substituting  our three estimates into (\ref{eq:sigma}) yields $\sigma_{\Vnorm} / \Vnorm \simeq 1.3$.  This translates to a standard deviation in log space of unity. 
% sqrt(0.95^2 + 0.8^2 + 0.19*0.6^2) = 1.3

% n Figure \ref{fig:elevations} we can see substantial variation in the amplitudes of elevation profiles among rocky bodies in the solar system. Here we have used the rms elevation profiles presented in \cite{lorenz2011hypsometry} and  \citet{2016LPI....47.2959B}.  The value of $\sigma_{\Vnorm}$ could therefore be substantial because there are so many sources of variability across the population of habitable planets, none of which are expected to be negligible. 

\section{Results}  \label{sec:results}
% , and here we make use of the  fiducial model outlined in the previous sectio
Our results are presented in three parts. The first part is largely pedagogical, where we use our fiducial model to illustrate the kind of selection effects that could arise among a fixed population of planets.  In the second part we shall allow this fiducial model to vary, in order to identify whether there is empirical evidence for a selection effect.  Finally, in the third part, we shall infer whether other life-bearing planets are likely to be more or less ocean dominated than the Earth. 
 % explore the impact of changing the mean surface water volume among habitable planets, and 
 
% using a distribution of planetary water compositions motivated from the numerical simulations of \citet{raymond2007high}. The objective is to illustrate the observational selection effect that the Earth's water mass fraction may have been subject to.

\subsection{Oceanic Selection Bias}
 
In the left hand panel of Figure \ref{fig:bias} we illustrate a fiducial model for the distribution of habitable radii and water compositions  $p(R, \WMF)$. This corresponds to a median water mass fraction $\mu = 9 \times 10^{-3}$, a standard deviation in $\log W$  of  0.8,  and we fix the rms elevation profile to be $\stdElevation = 2.51 \, \mathrm{km}$. (One could also introduce scatter in the planet-to-planet value of $\stdElevation$, but in terms of the surface conditions this is equivalent to broadening the variance in $W$.) The solid and dashed contours represent the $68\%$ and $95\%$ confidence limits respectively. The Earth appears as an outlier in this case, with a drier composition than over $97\%$ of the ensemble of water-bearing planets.

Many of these planets are heavily dominated by water. This is reflected by the central panel of Figure \ref{fig:bias}, which shows how the mean habitable area $H$ varies across the two-dimensional parameter space.  The numerical values are given in units of the Earth's land area. While larger planets boast greater surface areas, they are more susceptible to immersion due to their enhanced water volume. This is responsible for the sloped angle of the shaded region. At very low water compositions, the habitable area is seen to diminish. This is due to the assertion that planets with very small oceans are likely to lose much of their habitable surface to desert.  

In accordance with equation (\ref{eq:habitability}),  the habitable area influences the fecundity of a planet, and hence the likelihood of observing the set of planetary parameters. The distribution of planets \emph{as sampled by observers} is presented in the right hand panel of Figure \ref{fig:bias}. These $68$ and $95\%$ confidence limits are derived from equation (\ref{eq:bayes}). The strong preference for lower values of $\WMF$, relative to the true ensemble, is associated with their greater available land area. The median water mass fraction in this panel is more than a factor of 20 lower that the median value in the left panel. The Earth no longer appears as an outlier, now that evolutionary selection effects have been accounted for.  

%  Water bias: 0.035633; Radius bias:1.1558
%  Median Ocean Surface Fraction: 0.99692; Observed Ocean Frac:0.7052

\subsection{Bayesian Model Selection}    \label{sec:models}

Figure \ref{fig:obsoceanpdf} explores the distribution of the oceanic coverage for a range of values of  $\mu$, the median value of the water mass fraction $\WMF$. In effect, each line corresponds to a single model, such as the one illustrated in Figure \ref{fig:bias}, and reflects the probability of finding a given ocean coverage, when choosing a planet at random. The same range of planetary radii is used as before.  Starting at the driest cases of $10^{-5}$ and $10^{-4}$, we see that most planets have surfaces dominated by land. Then we have a `goldilocks' value of  $10^{-3}$ which generates a relatively even balance in surface compositions. But higher values of $\mu$ quickly lead to a preponderance of waterworlds. This, like Figure \ref{fig:landfrac}, depicts the fine-tuning problem associated with the Earth's water content.  Unless the model is very close to the terrestrial value $(0.1\%)$  then the vast majority of habitable planets are either extremely dry or almost completely covered by water. 

A possible resolution to the fine-tuning problem can be seen in right hand panel of Figure \ref{fig:obsoceanpdf}. These are the same set of five models shown in  the left hand panel, but here we plot the probability distribution for an observer's host planet, $p_\oplus(\oceanfrac)$. The distributions in the two different panels are related by the fecundity, as shown in equation (\ref{eq:bayeshabitable}). The selection effect acts to regulate the land-ocean divide, since waterworlds and desert worlds are deemed unconducive for the formation and proliferation of land-based intelligent life. In particular, given our observed ocean coverage of $71\%$, those models with a water mass fraction higher than the Earth are associated with higher likelihood values.  The maximum likelihood value for  $\mu$ is approximately $1\%$, an order of magnitude greater than that of the Earth, yet consistent with the findings of \citet{raymond2007high}.  

Does Figure \ref{fig:obsoceanpdf} lead us to favour the selection hypothesis  $\Hs$ (right hand panel) over the null hypothesis $\Ho$ (left hand panel)? Is the Earth's ocean coverage evidence of anthropic selection? To proceed, we compute the Bayes factor $K$, 

\[
K = \frac{P(D | \Hs)}{P(D|\Ho)} = \frac{\int P(\mu |  \Hs) P(D | \mu, \Hs) \ud \mu }{\int P(\mu | \Ho) P(D | \mu, \Ho) \ud \mu} \, .
\]
Note that the likelihood of $\mu$, denoted $P(D | \mu)$, is represented by the height of the curves in Figure \ref{fig:obsoceanpdf}  at the point where they cross the vertical dashed line. For the vanilla model $M$, the likelihood of $\mu$ is confined to a narrow window. By contrast, the  model which invokes anthropic selection requires less fine tuning. This is reflected in the Bayes factor $K = 6.3$, which constitutes evidence in favour of anthropic selection which is classified as `substantial'. 

These findings are robust to a number of changes in the model. For example, the quoted $K = 6.3$ relates to $\alpha=1$, but for all three values under consideration, we consistently find $K>5$. Further modifications to the model are explored in \S \ref{sec:appendix}. Ultimately all that matters is the one-dimensional distribution $p(\Vnorm)$, and the only critical criterion is that $\sigma_{\Vnorm}/\Vnorm$ is of the order unity, or greater. This seems a reasonable assumption given the many different variables which feed into it.
% This may be thought of as Occam's razor, which is an intrinsic component of Bayesian reasoning. 

Further explanation and examples of hypothesis testing can be found in \citet{mackay2003information}.

\begin{figure*} 
\includegraphics[width=80mm]{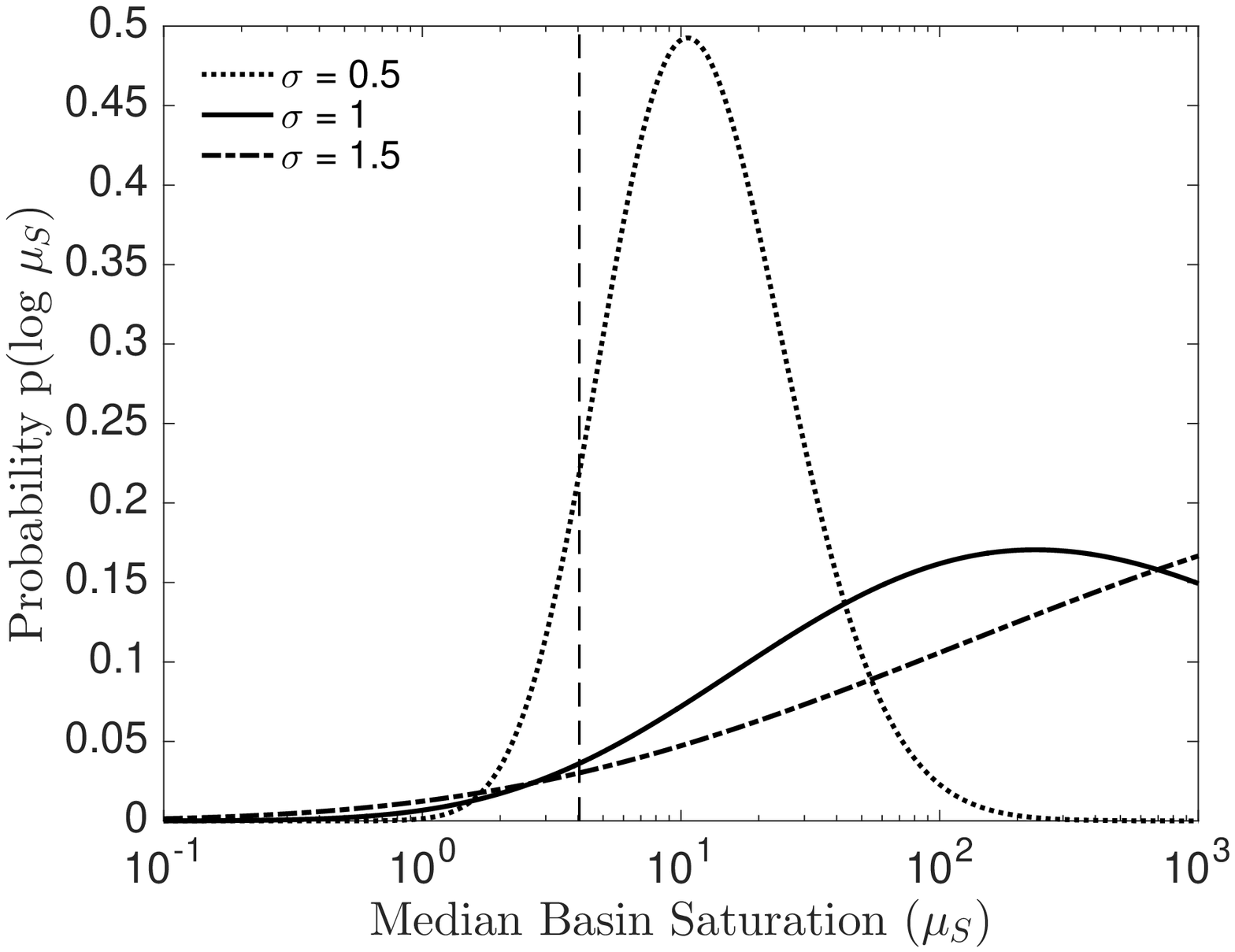} 
\hspace{0.2cm}
\includegraphics[width=80mm]{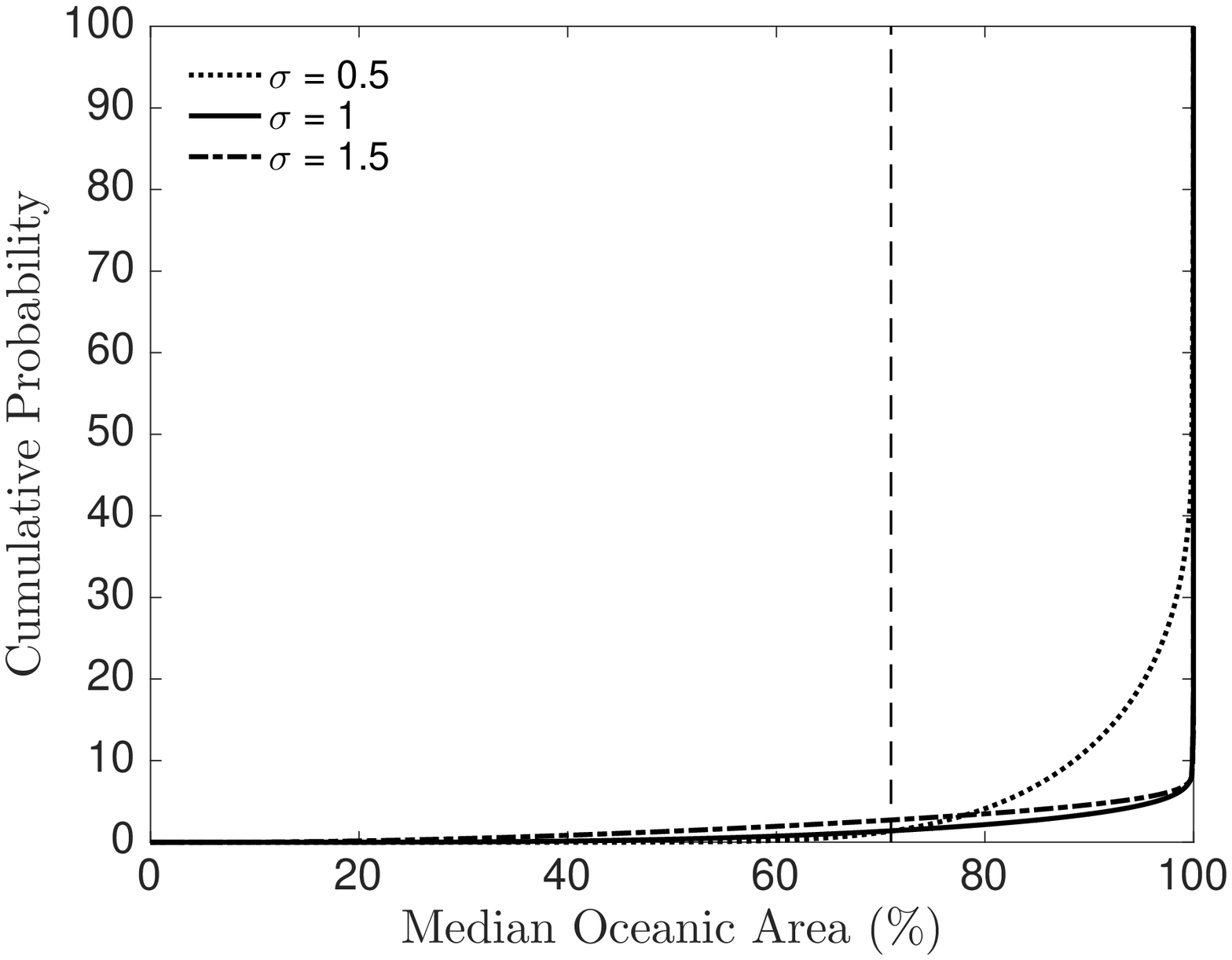} 
\caption{The posterior probability distributions for the median surface water characteristics among habitable planets. In each case the terrestrial value is marked as a vertical dashed line. 
\emph{Left:} 
The median saturation $\Vnorm$, as defined in (\ref{eq:X}), among habitable planets. 
\emph{Right:} 
The  same analysis as the left hand panel, but now recast in terms of the median ocean coverage. For each value of $\sigma_{\Vnorm}$ under consideration, we find that the majority of habitable planets are dominated by oceans  ($98\%$ credible interval). 
}
\label{fig:posterior}
 \end{figure*}

\subsection{Is the Earth wet or dry?} 
% anticipated
Now we turn our attention to what may be inferred about the ensemble of other life-bearing worlds. First we need a framework for the basin saturation values $\Vnorm$. Motivated by the anticipated scatter from the contributing factors in (\ref{eq:saturation}), we consider a Gaussian distribution in  $\log \Vnorm$, such that $\log \Vnorm \sim \mathcal{N}(\mS, \sigmaX^2)$.  In order to infer the median value $\mS$ among habitable planets - and hence the typical expanse of exoplanetary oceans -  we shall evaluate the posterior probability $p(\mS | D)$, which may be derived from the likelihood  as follows
\[
p(\mS | D) \propto \int p(D | \mS) p(\mS) \ud \mS \, .
\]
We adopt an uninformative prior $p(\mS) \propto \mS^{-1}$ across the range $10^{-4} <  \mS < 10^{4}$.  The second important variable is $\sigmaX$. While it is challenging to predict the magnitude of $\sigmaX$, it is unlikely to be small, given the multitude of sources which contribute to its variation. These include (a) the amplitude of surface perturbations, (b) variable water abundances across stellar nebulae, (c) varying water mass fractions accumulated during the process of planetary formation, (d) variations in the sizes of habitable planets, and (e) variable rates of water loss. 

The left hand panel of Figure \ref{fig:posterior} illustrates the posterior probability  $p(\mS | D)$ for three different values of $\sigmaX$, and for reference we mark the terrestrial value $\Searth$ with a vertical dashed line.  There is a clear preference for $\mS > \Searth$,  and that this preference strengthens for greater values of $\sigmaX$.  For our fiducial variance  in $\Vnorm$ $(\sigmaX = 1)$, we find $p(\mS > \Searth) = 0.91$. Very small values of $\sigmaX$ would suggest an extremely limited range of surface conditions among the ensemble of habitable planets, and so by construction we would expect our planet to be representative of others, irrespective of selection effects. 
 
The right hand panel of Figure \ref{fig:posterior} depicts the cumulative probability of the median oceanic coverage, and as with the left hand panel, we explore three different values for $\sigmaX$. For our fiducial value, $\sigmaX = 1$, we find that most are heavily water dominated $(\oceanfrac>90\%)$  ($95\%$ credible interval). Meanwhile our confidence that most are water dominated $(\oceanfrac>50\%)$ exceeds $99\%$.

These results use the habitable land area defined in (\ref{eq:H}), with $\alpha=1$. Lower values of $\alpha$ are associated with stronger confidence, while higher values of $\alpha$ weaken the conclusion. Yet even for the case $\alpha=5$, which is associated with an extremely rapid onset of desertification (see Figure \ref{fig:desert}), our confidence that the Earth is relatively dry remains over $80\%$. 

Why should we favour a scenario which actually deviates from our single observational value?  This ultimately stems from the highly nonlinear relationship between the water volume and the resulting ocean coverage.  If most habitable planets are waterworlds, then those few planets with some exposed continents will tend to be ocean-dominated. Conversely, if habitable planets tend to be land-dominated, there is little reason to believe an observer should find themselves in the narrow window of parameter space that produces an ocean-dominated planet.  This conclusion is not dependent on the details of our chosen model - it will arise for \emph{any} function $f(\Vnorm)$, provided $\sigmaX$ is not very small. Further evidence to support the robustness of our model is presented in Appendix \ref{sec:appendix}.

% \subsection{The source of bias} 

\section{Conclusions} \label{sec:conclusions}
On a purely statistical basis, one na\"{\i}vely expects to find a highly asymmetric division of land and ocean surface areas. A natural explanation for the Earth's equitably partitioned surface is an evolutionary selection effect. We have highlighted two mechanisms which could be responsible for driving this selection effect. First of all, planets with highly asymmetric surfaces (desert worlds or waterworlds) are likely to produce intelligent land-based species at a much lower rate. Secondly, planets with larger habitable areas are capable of sustaining larger populations. Both of these factors imply that our host planet has a greater habitable area than most life-bearing worlds.

% Key conclusions
We have exploited this model of planetary fecundity to draw two major conclusions. First of all, we find that the Earth's oceanic area provides substantial   evidence in favour of the selection model. Secondly, in the context of this model, we find that most habitable planets have surfaces which are over $90\%$ water ($95\%$ credible interval). Our results are robust to a broad variety of modifications to the model. The only critical assumption is that there is a significant variance in the basin saturation among habitable worlds, specifically $\sigmaX/\Vnorm \gtrsim 0.5$.  This appears likely given that there are many variable factors which contribute to a planet's surface water volume and basin capacity. 

The anticipated prevalence of waterworlds is driven by the fact that our home planet is close to the waterworld limit. Such proximity to a critical limit is precisely what one expects to find in the presence of a selection effect, provided only a smooth tail of the distribution lies below the critical limit. This reasoning was previously exploited by \citet{1987PhRvL..59.2607W} to successfully predict the value of the cosmological constant. 

% Implications
If the Earth's basin saturation is biased low, this implies that (a) its water mass fraction is likely to be biased low and (b) its elevation amplitude is likely to be biased high (and as with the basin saturation, the magnitude of this bias will depend on the planet-to-planet variance of these quantities). Do these two scenarios appear feasible? The water mass fraction among habitable planets could be considerably higher than the Earth. For example, numerical simulations based on delivering water from planetary embryos found a median water mass fractions of approximately $1\%$  \citep{raymond2007high}, ten times higher than the terrestrial value.  Extremely elevated water compositions have been associated with the inflation of planetary radii \citep{2016Thomas}. This scenario, in which the Earth is among the driest habitable planets, could help explain the appearance of a low-mass transition in the mass-radius relation of exoplanets \citep{2015Rogers, 2016ChenKipping}.
% In this work we have demonstrated that the Earth's water composition and ocean coverage may be typical among observers, yet highly atypical among life-bearing planets. 

If it transpires that the Earth is indeed unusually dry for a habitable planet, then one might wonder what the mechanism was. Does the Solar System have some distinguishing feature that was responsible? For example, perhaps the low eccentricities and inclinations of solar system planets are inefficient at promoting water delivery. Another possibility could be the influence of the Grand Tack model, where Jupiter underwent a reversal of its migration \citep{walsh2011low}. This has been found to yield a delivery of water that is approximately consistent with terrestrial levels  \citep{2014Icar..239...74O}. However recent simulations of the Grand Tack scenario suggest that, if anything, this may enhance  the delivery of water to terrestrial planets \citep{2016Matsumura}, rather than curtail it. Alternatively, a dry Earth may not necessarily have arisen from an identifiable macroscopic feature, it could simply be associated with the inherently stochastic nature of the water delivery process.

It also appears feasible that the Earth has an unusually deep ocean basin. The gravitational potential associated with its surface fluctuations is much higher than any other body in the solar system.  In turn this may suggest the Earth has unusually strong tectonic activity, and consequentially, an abnormally strong magnetic field. This exemplifies how selection effects can easily be transferred to correlated variables.  

Could the planet-to-planet variability in $\Vnorm$ be very small? Feedback mechanisms may have acted to regulate the depths of planetary oceans relative to the magnitude of their surface perturbations \citep{2012Abbot}. Earlier we denoted this possibility $\mathcal{H}_2$. Fortunately this hypothesis leads to a very different forecast for the surface conditions of Earth-like planets. If $\mathcal{H}_2$ is correct, we shall discover that a substantial proportion of habitable planets share the Earth's equitable water-land divide. This is in stark contrast to the prediction of our selection model, based on $\mathcal{H}_1$, where habitable planets are dominated by oceans.  

Other aspects of the Earth's surface that are susceptible to selection effects include the spatial configuration of land. For example, if a planet's land area were retained in a single contiguous piece, akin to Pangea, it may be that either a larger proportion of the land is rendered uninhabitable, or the ecological diversity is significantly suppressed.  The Earth's land configuration may be optimised to ensure that the majority of the available area is habitable, thereby maximising its fecundity, as defined in (\ref{eq:bayeshabitable}). 

This work builds on \citet{2015SimpsonAliens} by providing a further demonstration of why the Earth is likely to appear as a statistical outlier, across a broad spectrum of physical properties, when compared to other life-bearing worlds. In general, if a planet's population size is correlated with any variable, then the mean value witnessed by individuals will always exceed the true mean.  This is true for \emph{any} distribution of population sizes (see Appendix \ref{sec:proof}). 

% By modelling the ensemble of surface conditions, and  utilising the Earth's ocean coverage of $71\%$, our results suggest that the surfaces of most life-bearing planets are heavily water-dominated ($>99\%$ ocean coverage).  
% % It also does not discount the existence of water-based observers.
% Establishing a selection process based on land-based species does not discount the plausibility of water-based observers. There may be a number of water-based and land based observers, but \emph{a priori} it is extremely unlikely that these two numbers are a similar order of magnitude. And since we find ourselves to be land-based observers, it is highly probable that we are vastly more numerous than any water-based counterparts. 

%  These results are valid when adopting either the self-sampling assumption  \citep{bostrom2002self} or the self-indication assumption \citep{bostrom2003doomsday}, because the ensemble is local.

% Stats
 Ordinarily a single random sample is not particularly helpful in informing us on the nature of a broad population. However this limitation only applies to \emph{fair} samples. A single \emph{biased} sample can be used to place a lower (or upper) bound on the entire population distribution. For example, if the only data point we had regarding human running speed was taken from an Olympic 100 metres final, then we can be confident that a subsequent fair sample, across the global population, would not be significantly faster. Provided the population variance is significant, than we can be confident in finding a substantial deviation between the fair sample and the biased sample.
 
To give a further pedagogical example: imagine that you look at a kitchen worktop and notice some spilled coffee granules. One of those granules, selected at random, is found to lie within $0.1mm$ from the edge of the $600mm$ worktop. This proximity could of course be entirely coincidental. But it is much more likely that the bulk of the granules fell on the floor, and what you are seeing is merely the tail end of the distribution. 
  
The fine-tuning of the Earth's parameters is closely related to the proposition that various cosmological parameters correspond to those which optimise star formation \citep{tegmark2006dimensionless}. The key difference here is that many elements in the planetary ensemble are observable, and thus our predictions are experimentally falsifiable.  Indeed, it may not be long before we begin to build a census of nearby habitable planets, and begin to develop an understanding of how the Earth compares to other habitable worlds \citep{catala2009plato}.  If habitable planets systematically differ from the Earth in some way - such as the ocean coverage discussed in this work - this provides a hint as to the conditions which favoured the development of intelligent life. It would show that there is a bias between the inner sets of Figure \ref{fig:sets}. This bias tells us something about why we evolved on this particular lump of rock.
 
It has been argued that the finely-tuned properties of our planet is indicative of the sparsity of life in the Universe - the so-called `Rare Earth hypothesis'. However this interpretation overlooks one of the key factors which control the selection effect: the number of observers produced by each planet. The conditions on an individual's home planet is heavily skewed in favour of those conditions which maximise the \emph{abundance} of life. As an analogy, consider the contiguous piece of dry land you live on. It \emph{is} extremely special, in the sense that it is one of the largest pieces of contiguous land on the Earth's surface. But at the same time, there are hundreds of thousands of smaller chunks of land scattered across the Earth's surface. The selection effect that takes place when studying the ground beneath your feet is not a fair one.  Likewise, the rarity of the Earth's parameters need not reflect the sparsity of life in the cosmos. On the contrary, it may be driven precisely because we are a small piece within a vast ensemble.

When physiologists seek a deeper understanding of our body's features, such as our eyes and ears, a great deal of progress can be made from laboratory experimentation. Yet the \emph{only} way to arrive at a comprehensive answer is by including a complementary analysis of our origins. This allows biological function to be placed in an evolutionary context. A similar statement can be made regarding the features of our planet.  No matter how formidable our understanding of planet formation becomes, one can never hope to fully appreciate the Earth's features without addressing the issue of how we came into being upon it.

 \section*{Acknowledgements}
The author would like to thank Sandeep Hothi, Re'em Sari, Nick Cowan, Dorian Abbot, Eric Lopez, and the anonymous referees for helpful comments.  The author acknowledges support by Spanish Mineco grant AYA2014-58747-P and  MDM-2014-0369 of ICCUB (Unidad de Excelencia `Mar\'ia de Maeztu').

\appendix

\section{ Robustness of  the Results } \label{sec:appendix}

\begin{figure}
\includegraphics[width=80mm]{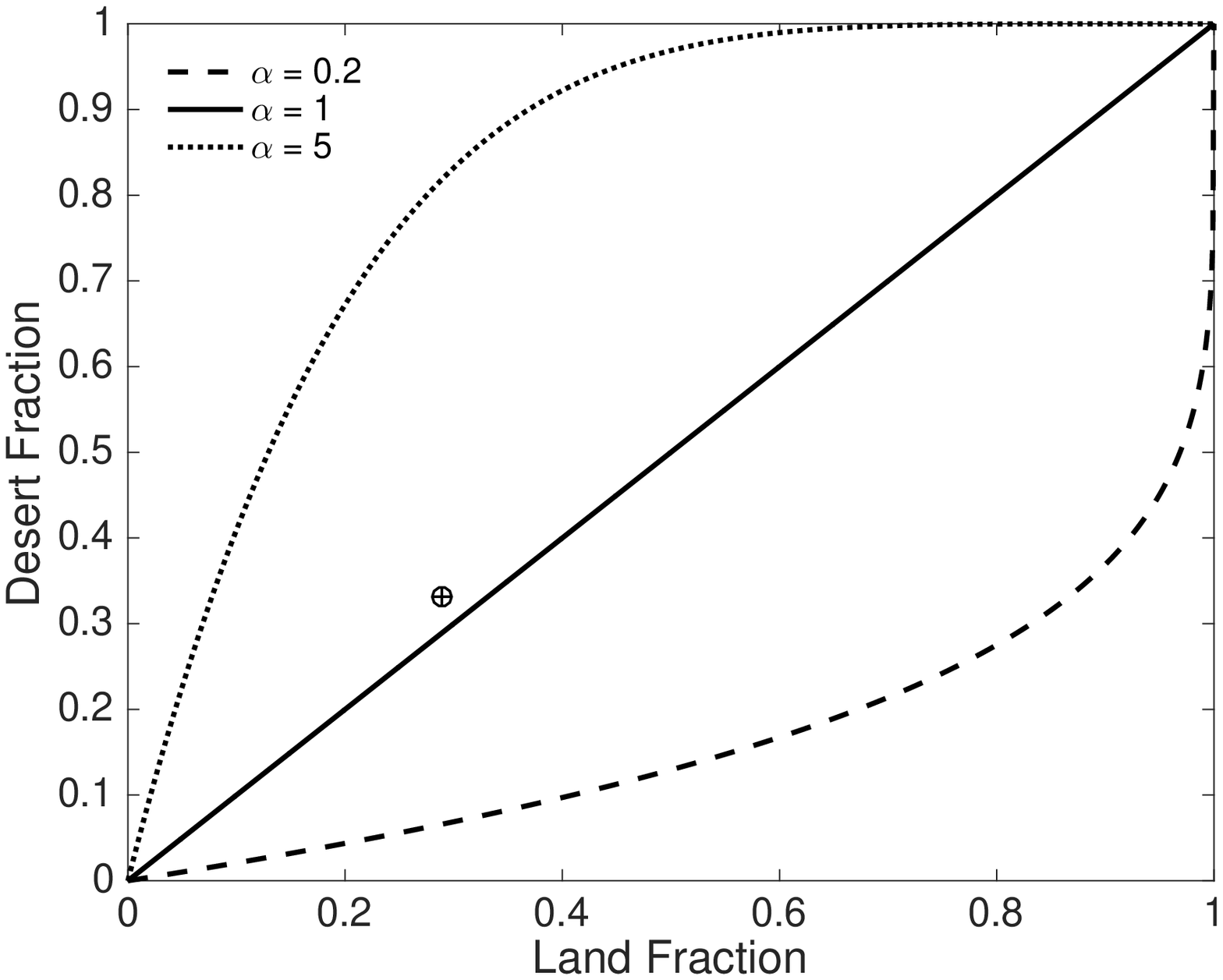}
\caption{The three different models of how the mean uninhabitable area approaches the full desert state associated with a dry planet. The solid line reflects our fiducial model ($\alpha = 1$), while the lower and  upper dashed lines represent the cases where $\alpha = \frac{1}{5}$ and $\alpha = 5$ respectively. These three models are used in Figures \ref{fig:alphavals} and \ref{fig:obsoceanpdf}.  
\label{fig:desert}}
\end{figure}

\begin{figure*}
\includegraphics[width=80mm]{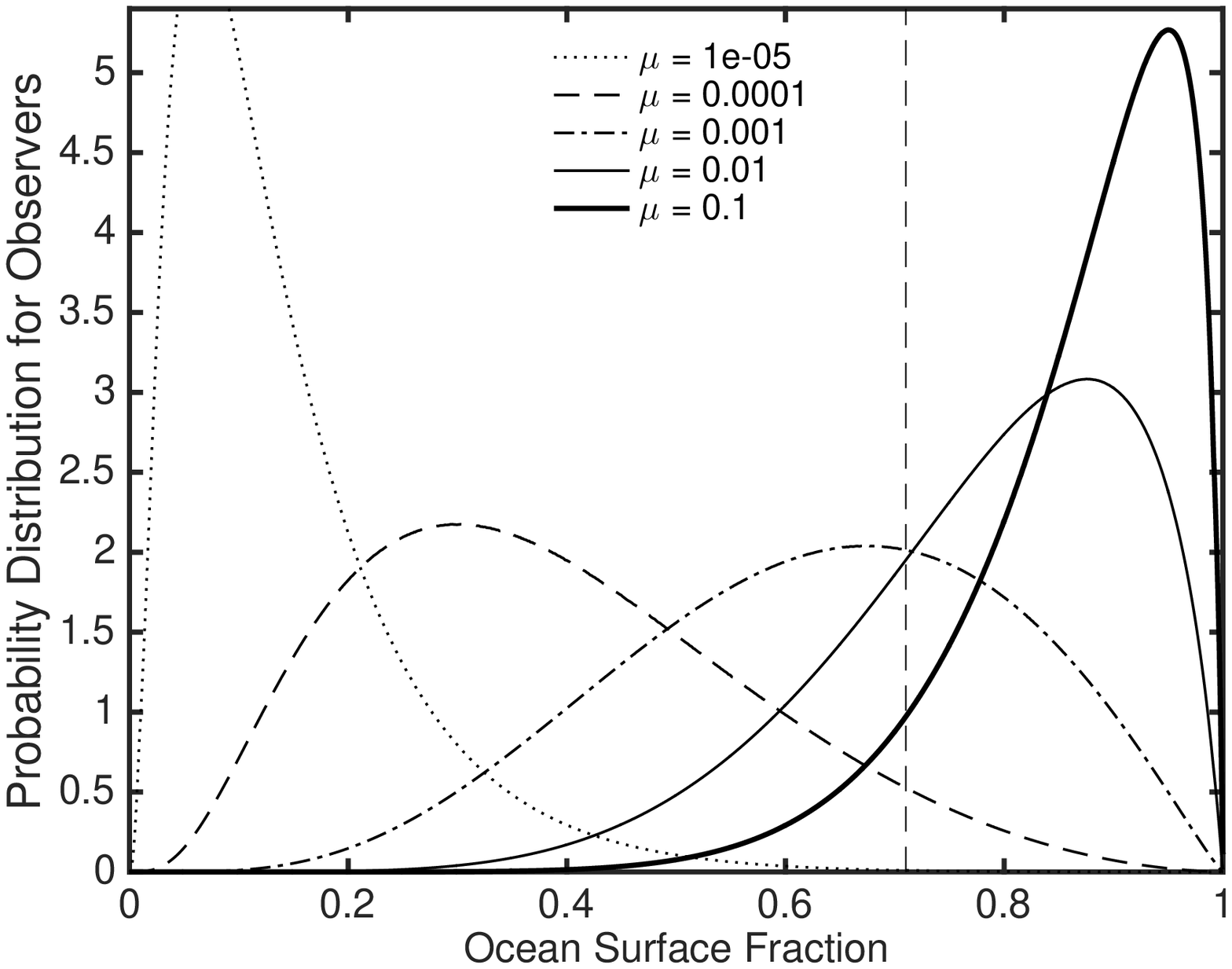}
\hspace{0.2cm}
\includegraphics[width=80mm]{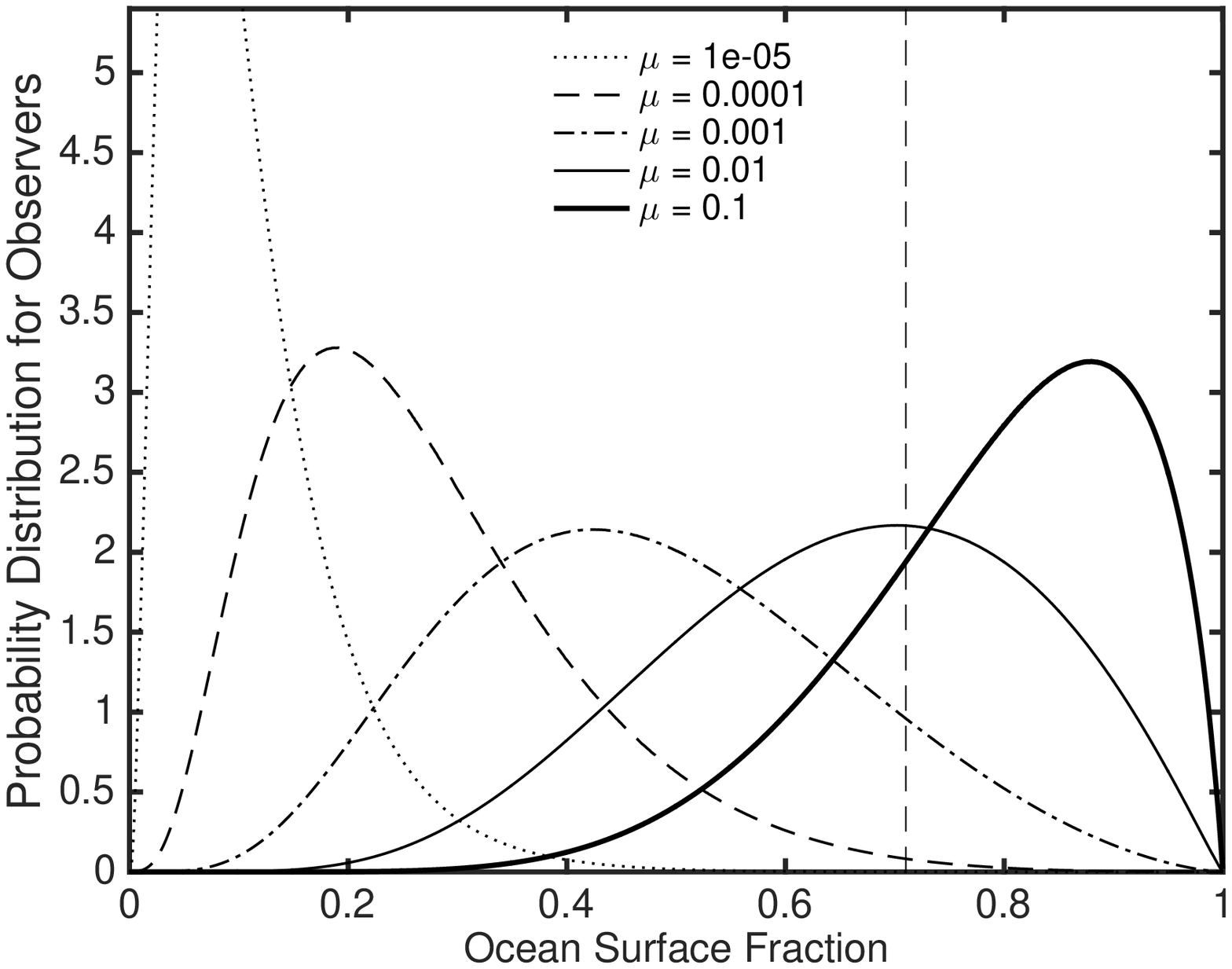}
\caption{Each panel is in the same format as the right hand panel of Figure \ref{fig:obsoceanpdf}, but here we alter the values of $\alpha$ which controls how the mean habitable area evolves as a function of ocean coverage, as specified by equation \ref{eq:desert}. In the left panel we set $\alpha = \frac{1}{5}$, while in the right panel $\alpha = 5$.}
\label{fig:alphavals}
\end{figure*}

\begin{figure*}
\includegraphics[width=200mm]{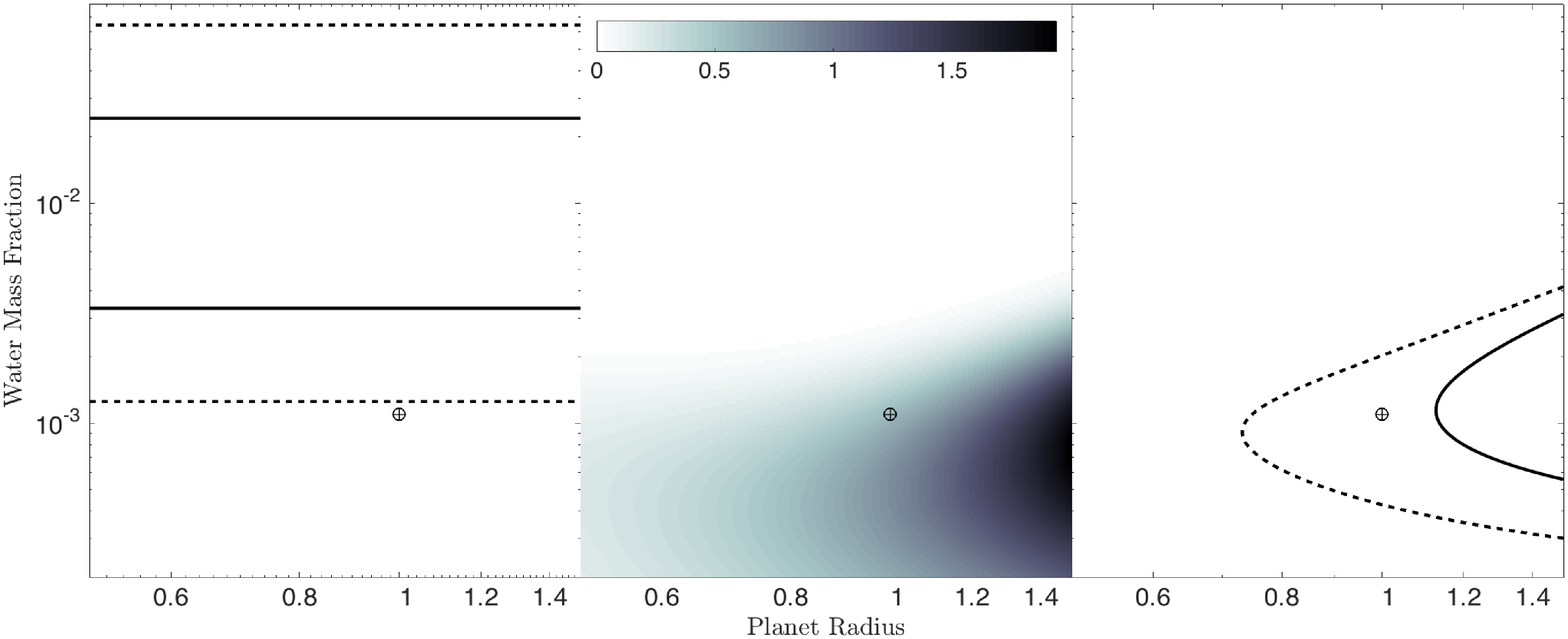}
\caption{The same format as Figure \ref{fig:bias}, but adopting a deep water cycle which leads to larger planets holding a greater proportion of their water content within the mantle. The radial dependence is modified, but the overall effect of suppressing the observed water composition remains unchanged.}
\label{fig:cycle}
\end{figure*}

\begin{figure*}
\includegraphics[width=80mm]{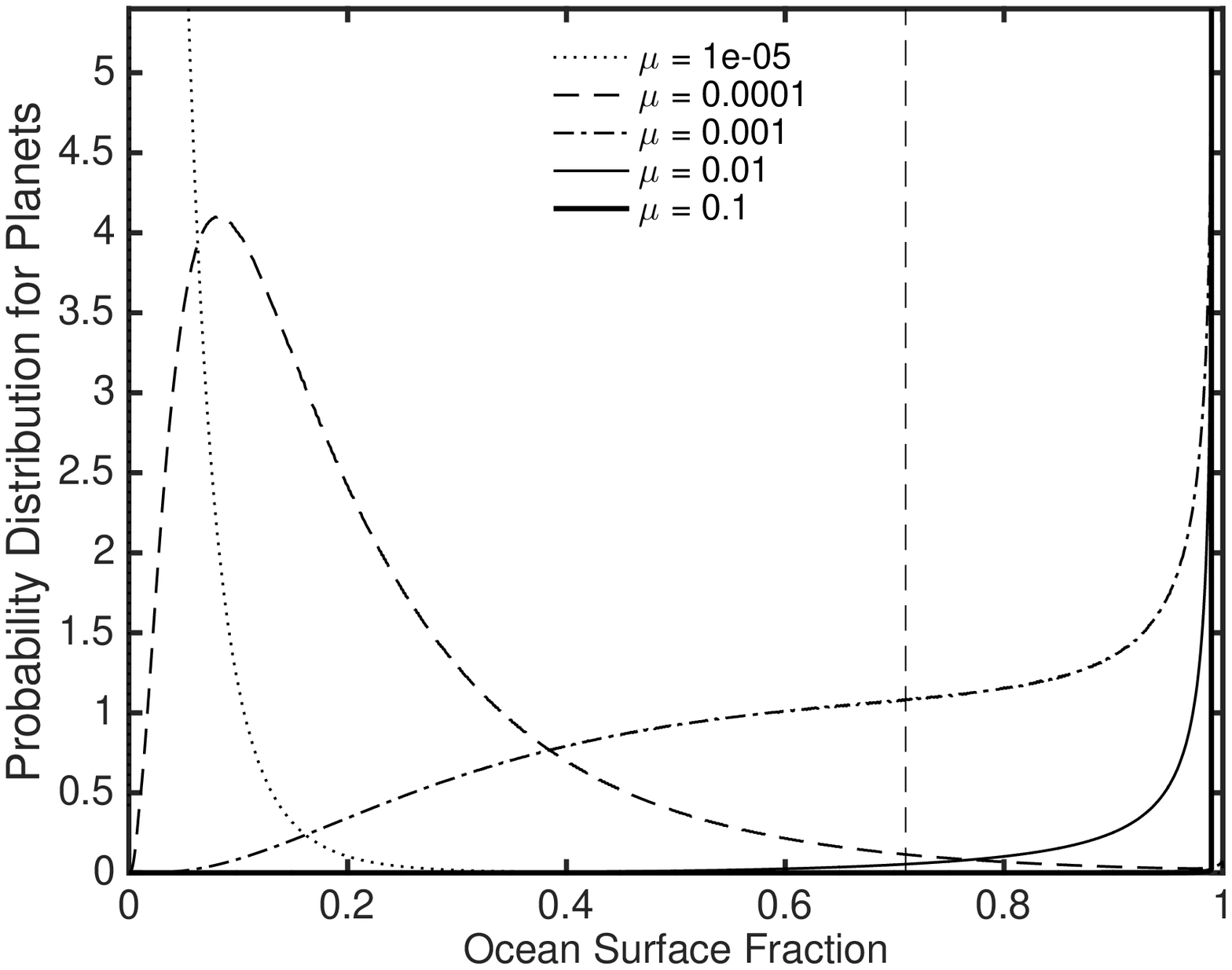}
\hspace{0.2cm}
\includegraphics[width=80mm]{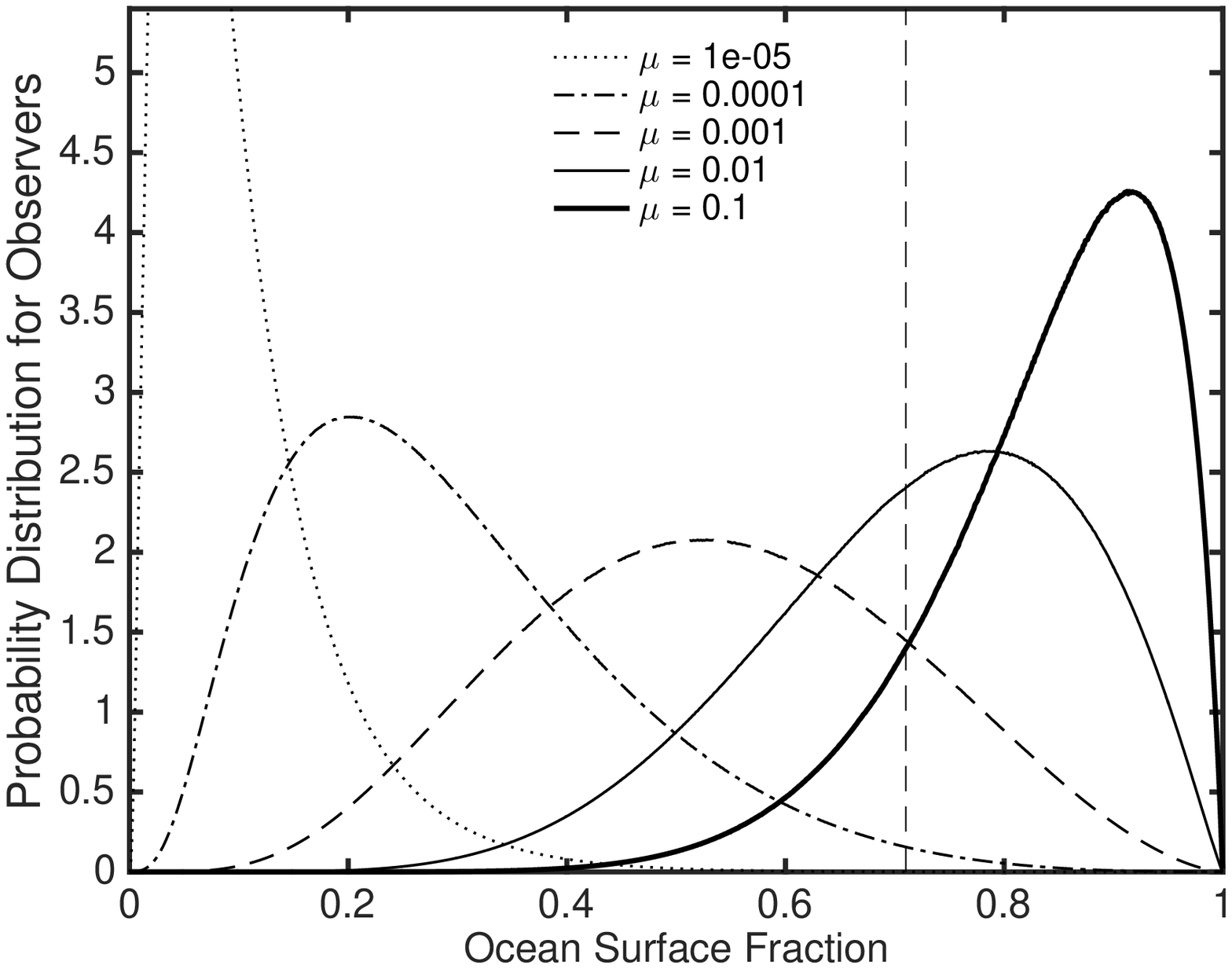}
\caption{The same format as Figure \ref{fig:obsoceanpdf}, but here we adopt a feedback model where the mantle water fraction evolves in accordance with \citet{cowan2014water}. }
\label{fig:cowan}
\end{figure*}

In this section we explore the impact of deviations from the fiducial model, first in terms of the habitable land area $H$, and then the fraction of water stored on the surface $f_s$. 

\subsection{Modelling the habitable land area}

The fiducial model assumed a linear progression from a desert-dominated landmass, to a fully habitable one. This was defined in equation (\ref{eq:desert}), taking $\alpha = 1$. Clearly in reality the progression may take on a different functional form, so here we shall explore the influence of the exponent $\alpha$. 

In Figure \ref{fig:desert} we can see the power law relations which we use to model the fraction of land which is rendered an uninhabitable desert. The solid line represents the fiducial case ($\alpha = 1$), while the lower dashed line illustrates $\alpha = 5$. The upper dashed line corresponds to $\alpha = \frac{1}{5}$, and this would correspond to a mean fraction of land lost to desert of over $75\%$, for planets with an Earth-like ocean coverage. 

There is of course some ambiguity in the point at which land becomes uninhabitable.  As shown by the datapoint in Figure \ref{fig:desert}, some $33\%$ of the Earth's land is classified as desert. A more stringent classification is a region which receives less than 250mm of rainfall per year, which severely compromises the ability for macroscopic organisms to thrive.  Approximately $14\%$ of the Earth's land mass satisfy this criteria \citep{cordey2013north}.

Figure  \ref{fig:alphavals}  demonstrates the impact of  $\alpha$ on the likelihood of different planetary water compositions.  The left panel adopts $\alpha = \frac{1}{5}$, yet still bears a close resemblance to the fiducial case of  $\alpha = 1$. The drier configurations are now more viable, since desert form more aggressively in this model. 
The right panel show the result for  $\alpha = 5$, and again the trend is much the same, closely resembling the result of Figure \ref{fig:obsoceanpdf}.   The tendency for the data to prefer water-rich models therefore holds for a broad range of desert models.

\subsection{Modulating the surface water fraction}

In our fiducial model we assume that the proportion of a planet's water which resides on the surface, $f_s$, does not systematically change with planetary radius or water composition. \citet{cowan2014water} present a model for a variable mantle water fraction, which scales with the planet's surface gravity.  In Figure \ref{fig:cycle} we repeat the procedure used to produce Figure \ref{fig:bias}, but now employ the scaling relation which leads to more massive planets retaining a greater proportion of their water inventory within the mantle. This enables them to posses large areas of habitable area, as is apparent from the central panel. Nonetheless the key outcome in the right panel, a strong reduction in the observed water composition, is unchanged from the fiducial model. 

In Figure \ref{fig:cowan} we see the impact the feedback model  of \citet{cowan2014water}  has on the ocean coverage distribution, both for the total ensemble (left panel) and the ensemble of observers (right panel). In this feedback model, planets less massive than the Earth retain less water in the mantle, and so are more prone to flooding. Conversely more massive retain more water in the mantle, so not as susceptible to flooding as they had been without the feedback model. Since we are viewing the average over the full ensemble, the net effect is barely distinguishable from the fiducial model of Figure \ref{fig:obsoceanpdf}. 

Note that this model still does not explore $f_s$ evolving as a function of $\WMF$, which would arise if a feedback mechanism operates to regulate the ocean volume. However as discussed in \S \ref{sec:feedback}, there are fairly stringent limits on the strength of this feedback effect, due to the finite capacity of the mantle.  

\section{Group Selection Bias} \label{sec:proof}

Here we present a brief proof that selecting an element at random (an observer in the context of this work) will always lead to an amplification of any quantity which is correlated with the population of the group. 

Given a set of elements $\set$ partitioned into $N$ subsets, we define  $n_i$ as  the number of elements in the $i$th subset. If each subset is assigned a value for a secondary variable $r_i$, and this variable $r$ is positively correlated with $n_i$, then by definition 

\[
\sum_i (r_i - \bar{r})(n_i - \bar{n}) > 0  \, ,
\]
or equivalently
\[
\sum_i n_i r_i - N \bar{r} \bar{n} > 0\, .
\]

This may be rearranged as follows

\[
\frac{\sum_i n_i r_i}{\sum_i \bar{n}} > \bar{r}\, ,
\]
where the expression on the left hand side is the definition of the elemental mean $\bar{r}_{e}$. The elemental mean therefore always exceeds the group mean, for any distribution $n_i$
\[
\bar{r}_{e} > \bar{r} \, .
\]

\bibliographystyle{mnras}
\bibliography{../anthropic_bib}

\bsp	 
\label{lastpage}
\end{document}